\documentclass[fleqn,usenatbib]{mnras}

\usepackage[flushleft]{threeparttable}
\usepackage{array}
\usepackage{mathtools}  
\usepackage{graphicx}
\usepackage{subfigure}
\usepackage{grffile}
\usepackage{eqparbox}
\usepackage{natbib}

\usepackage[T1]{fontenc}

\DeclareRobustCommand{\VAN}[3]{#2}
\let\VANthebibliography\thebibliography
\def\thebibliography{\DeclareRobustCommand{\VAN}[3]{##3}\VANthebibliography}

\usepackage{graphicx}	
\usepackage{amsmath}	
\usepackage{amssymb}	
\usepackage[normalem]{ulem}
\usepackage{soul}
\usepackage{url}
\usepackage{orcidlink}

\def\msun{\mbox{M$_{\odot}$}} 
\def\mbcg{\mbox{$M_{\rm \ast,BCG}$}}
\def\micl{\mbox{$M_{\rm \ast,ICL}$}}

\def\mstartotal{\mbox{M$_{\rm \ast,200}$}}
\newcommand{\Msun}{{\rm M}_{\odot}}
\newcommand{\Mtwo}{M_{\rm 200}}
\newcommand{\Mfive}{M_{\rm 500}}
\newcommand{\Mtwost}{M_{\rm \ast, 200}}
\newcommand{\Rtwo}{R_{\rm 200}}
\newcommand{\Rfive}{R_{\rm 500}}
\newcommand{\ctwo}{c_{\rm 200}}
\newcommand{\zfive}{z_{\rm 50}}
\newcommand{\deltam}{\Delta M_{\ast,\rm 4th}}
\newcommand{\rhalf}{$r_{\rm \ast,half}$}
\newcommand{\rhalfpro}{$r_{\rm \ast,half}^{\rm 2D}$}

\newcommand{\ficl}{f_{\rm ICL}}
\newcommand{\ficltot}{F_{\rm ICL}}
\newcommand{\ficlpro}{f_{\rm ICL}^{\rm 2D}}
\newcommand{\ficltotpro}{F_{\rm ICL}^{\rm 2D}}

\defcitealias{montenegro2023}{MT+23}

\usepackage{graphicx}	
\usepackage{amsmath}	

\title[Stellar distribution of galaxy clusters in TNG300]{The stellar mass composition of galaxy clusters and dependencies on dark matter halo properties}

\author[D. Montenegro-Taborda et al.]
{
\parbox{18cm}{
Daniel Montenegro-Taborda$^{\orcidlink{0000-0002-5428-9984}}$,$^{1}$\thanks{E-mail: d.montenegro@irya.unam.mx}
Vladimir Avila-Reese$^{\orcidlink{0000-0002-3461-2342}}$,$^{2}$
Vicente Rodriguez-Gomez$^{\orcidlink{0000-0002-9495-0079}}$,$^{1}$\\
Aditya Manuwal$^{\orcidlink{0000-0003-2893-2793}}$,$^{2}$
and Bernardo Cervantes-Sodi$^{\orcidlink{0000-0002-2897-9121}}$$^{1}$
}
\vspace{0.3cm} \\ 
$^{1}$Universidad Nacional Aut\'onoma de M\'exico, Instituto de Radioastronom\'{\i}a y Astrof\'{\i}sica. 58089. Morelia, Michoac\'an, M\'exico\\
$^{2}$Universidad Nacional Aut\'onoma de M\'exico, Instituto de Astronom\'{\i}a. A.P. 70-264, 04510. Ciudad de M\'exico, México
}

\date{Accepted XXX. Received YYY; in original form ZZZ}

\pubyear{2024}

\begin{document}
\label{firstpage}
\pagerange{\pageref{firstpage}--\pageref{lastpage}}
\maketitle

\begin{abstract}
We analyze 700 clusters from the TNG300 hydrodynamical simulation ($\Mtwo\geq5\times10^{13} \,\Msun$ at \(z=0\)) to examine the radial stellar mass distribution of their central objects, consisting of the brightest cluster galaxy (BCG) and the intracluster light (ICL). The BCG+ICL mass fraction weakly anticorrelates with $\Mtwo$, but strongly correlates with the concentration, $c_{200}$, the assembly redshift, $z_{50}$, and the mass gap between the most massive and the fourth more massive member, $\deltam$. We explore different aperture radii to nominally separate the ICL from the BCG and calculate ICL fractions. For $r_{\rm{ap}}=2$\rhalf, where \rhalf\ is the radius containing half the BCG+ICL mass, the ICL fraction is nearly independent of $\Mtwo$, $c_{200}$, and $z_{50}$ with values $\micl/(\micl+\mbcg)= 0.33\pm0.03$. Including the stellar mass of the satellites, the fraction $\micl/(\micl+\mbcg+M_{\rm \ast,sat})$ weakly anticorrelates with $\Mtwo$ and strongly correlates with $c_{200}$, $z_{50}$, and $\deltam$, suggesting that in more concentrated/earlier assembled/more relaxed clusters more stellar mass is lost from the satellites (by tidal stripping, and mergers) in favour of the ICL and BCG. Indeed, we find that ex-situ stars dominate both in the BCG and ICL masses, with mergers contributing more to the BCG, while tidal stripping contributes more to the ICL. We find that the difference between the projected and 3D ICL fractions are only a few per cent and suggest using 2\rhalfpro\ to separate the ICL from the BCG in observed clusters.

\end{abstract}




\section{Introduction}
\label{sec: Intro}

Clusters of galaxies are the most massive bound structures in the Universe, with virial masses larger than $0.5-1 \times 10^{14}$ \msun. They contain more than $50$ galaxies, hot intracluster gas, dark matter, and a very low surface brightness diffuse light that normally surrounds the brightest central galaxy (BCG), located in the deepest potential well of the cluster. The diffuse component, called intracluster light (ICL) since it was predicted \citep{zwicky1937} and observed in the Coma cluster \citep{zwicky1951_ICL}, is produced by stars that are assumed not to be bound to any particular galaxy, but to the potential of the cluster itself. ICL structures are the most extended stellar systems in the Universe, with characteristic radii of hundreds of kiloparsecs up to the megaparsec scale. In recent years, the observational and theoretical study of the ICL has intensified rapidly thanks to the possibility of obtaining very deep observations of large samples of clusters, and to major advances in computational facilities and modelling of galaxy formation and evolution \citep[see for recent reviews, ][]{montes2022Nature,contini2021_Review}.

According to the hierarchical clustering scenario based on the $\Lambda$CDM cosmology \citep[]{Mo+2010}, the cluster dark matter haloes are built hierarchically by relatively late smooth accretion and merging of substructures; most clusters are still in their active assembly phases today. The luminous systems within the substructures also grow by accretion and mergers, and as they interact with each other and with the strong tidal field of the cluster, they lose stars (as well as gas). As a result, the central galaxy in the halo grows by in situ star formation from the accreted gas and by mergers, while the ICL forms from the stripped stars of surviving galaxies, from the stars ejected in mergers, and it is probable that also a small fraction by in situ star formation in the intracluster medium. The ICL can also grow by acquiring the diffuse stellar component in the infalling satellites and groups (preprocessing). All the above processes contribute different fractions to the mass of the BCG and ICL, depending on the time and mass regime \citep[see for recent reviews,][and the references therein]{Montes2019,contini2021_Review}. 

An interesting question is how much the BCG and ICL co-evolve.  The degree of co-evolution seems to be different at different epochs, and depends on the rate of cluster growth as well as its dynamical state \citep[e.g.,][]{Contini2018,Chun+2023}.
To explore theoretical results, in particular from cosmological hydrodynamics simulations, and to constrain different growth channels for BCG and ICL, we can study the dependences and correlations of their stellar mass fractions and spatial (3D) distributions with host cluster/halo properties. A crucial difficulty in this task is to bring these studies to the level of observations for direct comparisons. The problem is further complicated by the inherent difficulty in separating the BCG from the ICL, in particular when using photometric information in the case of observations. Several definitions and methodologies have been applied to this end, each one having its advantages and difficulties \citep[for recent discussions, see][]{kluge2021photometric,montes2022Nature,Brough+2024}. 

A first step towards determining suitable BCG-ICL separation criteria is to apply methods based on 3D information for stellar particles in simulations, and explore the \textit{intrinsic} dependences of the obtained ICL (and BCG) stellar mass fractions on the cluster/halo properties. This is the main goal of this paper. Next, one can investigate the extent to which these relations hold for projected (2D) distributions, and then for light. We discuss the former point here and show that our conclusions hold for projected quantities, while the latter will be investigated in detail in a companion paper. 

The first cosmological hydrodynamical simulations of galaxy clusters aided in discerning different channels of ICL formation, their contributions at different galactocentric radii, and the typical ICL fractions \citep{Murante+2004,murante2007,Willman+2004,Sommer-Larsen+2005}. 
Later studies based on simulations -- which significantly improved thanks to the vertiginous advance in computational capacities and the improvements in the subgrid physics schemes (e.g., the inclusion of AGN feedback) -- focused then on the details of which channels are most dominant on ICL production \citep[e.g.,][]{Puchwein+2010,Pillepich2018a,remus2021accreted,Tang+2023,montenegro2023}, how to define and determine the ICL component in a physical (dynamical) way \citep[e.g.,][]{dolag2010,Puchwein+2010,cui2014,remus2017,marini2022ML,contreras2024,proctor2024}, and how to compare the ICL fractions with those from optical observations \citep[][]{cui2014,Tang+2018, Henden2020,Brough+2024}.  
Alternative methods using dark matter-only cosmological simulations and empirical relations to insert galaxies \citep[]{rudick2006,rudick2007,rudick2011,cooper2015,Chun+2023}, or semi-analytical recipes to seed galaxies within the subhaloes and estimate the tidal stripping of stars \citep[]{contini2014formation,Contini2018,contini2019,contini2023}, have also contributed immensely to the understanding of these questions. 

A general consensus of most theoretical studies is that the ICL is a \textit{ubiquitous} stellar component whose inner density distribution is coupled to that of the BCG because the ex situ stars of both components overlap in a spatial region that can be relatively extended \citep[e.g.,][]{cooper2015,remus2017,Pillepich2018a,contini2022,montenegro2023,proctor2024}. 
This makes it very difficult to separate the two components on the basis of surface brightness distributions, although a \textit{nominal transition radius} could probably be defined where one component starts to clearly dominate over the other \citep[][]{Chen+2022,contini2022}. 
On the other hand, the definition of the ICL based on stellar particle dynamics methods is unfortunately not directly applicable to observational comparisons.
Importantly, simulations show that ICL fractions measured with these methods are significantly larger than those determined using different criteria based on the surface density or brightness profiles \citep{rudick2011,cui2014,Brough+2024}.

In this paper, following \citet[][hereafter   \citetalias{montenegro2023}]{montenegro2023}, we use 700 clusters ($M_{200}\geq 5\times 10^{13}\,{\rm M}_\odot$) from the IllustrisTNG (TNG300) simulation \citep{Marinacci2018,Naiman2018first,Nelson2018,Pillepich2018a, springel2018} to study the dependence of the BCG+ICL stellar mass fraction and BCG/ICL separation on several cluster properties. We focus on the $z=0$ 3D stellar mass distribution within the main stellar structure in the clusters (all stars contained in the main subhalo), and explore different radial apertures to separate the ICL from the BCG. Our goals are to find an optimal aperture that defines the transition radius, to study how the BCG+ICL and ICL mass fractions depend on the cluster dynamical state, and to find the cluster properties to which these mass fractions are most sensitive. We also study the distributions of the fractions of in situ and ex situ stars in the BCG and ICL, the contributions of different channels to the ex situ component, and their dependence on the cluster dynamical state.

We have organized this paper as follows. In Section~\ref{sec:methodology}  we detail the simulation, our sample, and definitions utilized in this study. In Section~\ref{sec:results} we present our main results of analyzing the radial stellar mass distribution for our 700 clusters and their dependency with specific cluster properties. The formation channels for the BCGs and ICL at $z=0$ are studied in Section~\ref{sec:chanels}. In Section~\ref{sec:discussion} we discuss our results and implications, and we compare them with previous work. Finally, in Section~\ref{sec:conclusions} we conclude the paper with a summary of our key findings.

\section{The methodology}
\label{sec:methodology}

\subsection{The TNG300 simulation}
\label{sec:maths} 

We use data from {\it The Next Generation Illustris Project\footnote{\url{www.tng-project.org}}} \citep[IllustrisTNG,][]{Marinacci2018, Naiman2018first, Nelson2018, Pillepich2018a, springel2018},  a suite of magnetohydrodynamic cosmological simulations run with the moving-mesh code \textsc{arepo} \citep{Springel2010-arepo}, which model the formation and evolution of galaxies within the $\Lambda$CDM paradigm. The cosmological parameters used are consistent with \citet{ade2016planck} measurements: $\Omega_\textrm{m}$ = 0.3089, $\Omega_{\Lambda}$ = 0.6911, $\Omega_\textrm{b}$ = 0.0486, $h$ = 0.6774, $\sigma_8$ = 0.8159, $n_\textrm{s}$ = 0.9667. The initial conditions were obtained at $z=127$ using the \textsc{N-GenIC} code \citep{springel2005simulations,springel2015n}, based on a linear-theory power spectrum.  IllustrisTNG features an updated version of the Illustris galaxy formation model \citep{vogelsberger2013model, torrey2014model} that includes gas radiative cooling, star formation and evolution, chemical enrichment, stellar feedback, and supermassive black hole (SMBH) growth and AGN feedback; described in \citet{Pillepich2018}.

In this work, we use the highest-resolution version of TNG300 (TNG300-1 in table 1 of \citealt{tngdata}), which follows the coevolution of dark matter (DM), gas, stars, and supermassive black holes (SMBHs) within a cubical volume of approximately 300 Mpc (comoving) per side, which models 2500$^3$ DM particles (resolution $m_{\rm DM}\approx 5.9\times10^{7} \msun$) and approximately 2500$^3$ baryonic elements (average mass resolution $m_{\rm b}\approx 1.1\times10^{7} \msun$; for more details, see \citealp[]{Springel2010-arepo}) constituted to gas cell or stellar particles. The gravitational softening length of DM and stellar particles is approximately 1.5 kpc at $z<1$, while the gas cells have an adaptive comoving softening with a minimum of 0.37 kpc. 

The simulation has 100 snapshots ranging from z$\sim$20 to z=0. For each of these snapshots, haloes and substructures (subhaloes) are identified by implementing two algorithms consecutively: friends-of-friends \citep[FoF;][]{davis1985evolution}  and \textsc{\large subfind} \citep{Springel2001-SUBFIND, dolag2009substructures}.
The FoF algorithm selects all the \textit{dark matter} particles separated by a distance less than or equal to 0.2 times the average particle separation, thereby identifying ``haloes'' within the simulation. The gas, stars and BH elements are attached to the haloes containing their nearest DM particles. 
The substructures are then identified through the \textsc{\large subfind} algorithm, which determines gravitationally bound overdensities within each halo by considering all the particle types. The catalogue generated by \textsc{\large subfind} contains subhaloes flagged as centrals or satellites. The position of the central subhalo coincides with the centre of its parent halo and determined by the location where the minimum gravitational potential is obtained; therefore, every halo has only one central subhalo. Furthermore, the central object not only contains the 
stellar particles constituting the central galaxy, it also contains
the intrahalo stars that are not bound to any satellite, i.e. the ICL. 

For this paper, we consider all TNG300 haloes with masses $\Mtwo \geq 5 \times 10^{13} \, {\rm M}_{\odot}$ at $z=0$, which yields a total of 700 objects, with 280 above $10^{14}$ \msun, and 3 above $10^{15}$ \msun. For comparison, the TNG100 simulation (of $\sim$100 Mpc per side) contains only 41 systems with $\Mtwo \geq 5 \times 10^{13} \, {\rm M}_{\odot}$ at $z=0$.

\begin{table*}
\begin{center}
\begin{threeparttable}
\begin{tabular}{ m{3.cm}  m{8cm} m{5cm} }
\hline
Name & Definition & Mass and fraction measurements \\ \hline
\\
Cluster of galaxies
&
\textit{} A spherical structure (containing DM, stellar and gas particles) of radius $R_\Delta$, defined as the radius where the spherical overdensity in a FoF group is $\Delta$ times $\rho_{\rm crit}$. Here, we use $\Delta=200$ and consider as clusters those with $\Mtwo \geq 5 \times 10^{13} \, {\rm{M}}_\odot$ at $z=0$.
&
$\Mtwo\equiv M(r<{\rm R_{200}})$ and, eventually, $\Mfive\equiv M(r<{\rm R_{500}})$ 
\\
\\
Total (cluster) stellar mass
& 
Total mass in stellar particles within the cluster (usually the virial radius), composed of the stellar particles within the main subhalo (central object) plus those within the other subhaloes (satellite galaxies).
&
$\Mtwost \equiv M_\ast({\rm r < \Rtwo})$
\\
\\
Central object or BCG+ICL & 
All stellar particles gravitationally bound to the main subhalo as identified by \textsc{\large subfind} and within the virial radius, that is, the stellar particles that are \textit{not locked} in satellite galaxies.\footnote{The so-called `fuzz', composed of particles that are not gravitationally bound to any substructure or to the halo as a whole, is negligible in the case of stars.} This corresponds to what we nominally call the central galaxy (BCG) plus the ICL.
&
$M_{\rm \ast, BCG+ICL} \equiv M_{\rm \ast, cen}\equiv M_{\ast}$(central object, $r<\Rtwo$)
\\
\\
BCG (brightest cluster galaxy)  & 
In this work, the BCG is defined by a radial cut in the central object's 3D stellar mass distribution at an aperture radius $r_{\rm ap}$, where $r_{\rm ap}$ can be a constant radius (30, 50, and 100 kpc) or a radius that depends on each cluster (0.05--0.1 $\Rtwo$ or 0.5--2\rhalf, where \rhalf\ is the stellar half-mass radius of the central object).
& 
$M_{\rm \ast, BCG} \equiv M_{\ast}$ (central object, $ r < r_{\rm ap} $)
\\
\\
ICL (intracluster light) & 
In this work, the ICL is defined as the complement to the BCG, that is, all stellar particles at $r>r_{\rm ap}$ and up to $\Rtwo$. 
&
$M_{\rm \ast, ICL} \equiv M_{\ast}$ (central object, $r_{\rm ap} < r < \Rtwo$) = $M_{\rm \ast, BCG+ICL} - M_{\rm \ast, BCG}$
\\
\\
Satellite galaxies
& 
All stellar particles in satellite galaxies  (subhaloes) within $\Rtwo$.
&
$M_{\rm \ast, sats} = \Mtwost- M_{\rm \ast, BCG+ICL}$
\\
\\
BCG+ICL mass fraction
& 
The BCG+ICL stellar mass fraction relative to the {\it total} stellar mass of the cluster (central object plus satellites) within $\Rtwo$. 
&
$F_{\rm BCG+ICL}\equiv \frac{M_{\rm \ast, BCG+ICL}}{M_{\rm \ast, BCG+ICL}+M_{\rm \ast, sats}}=\frac{M_{\rm \ast, cen}}{\Mtwost}$
\\
\\
Total ICL mass fraction
&
The ICL stellar mass fraction relative to the {\it total} stellar mass of the cluster (central object plus satellites) within $\Rtwo$.
&
$F_{\rm ICL}\equiv \frac{M_{\rm \ast, ICL}}{M_{\rm \ast,BCG+ICL}+M_{\rm \ast,sats}} = \frac{M_{\rm \ast,ICL}}{\Mtwost}$
\\
\\
ICL mass fraction
&
The ICL stellar mass fraction relative to the stellar mass of the {\it central} object (BCG+ICL) within $\Rtwo$.
&
$\ficl \equiv \frac{M_{\rm \ast, ICL}}{M_{\rm \ast, BCG+ICL}} = \frac{M_{\rm \ast, ICL}}{M_{\rm \ast,cen}}$
\\
\\
BCG mass fraction
&
The BCG stellar mass fraction relative to the stellar mass of the {\it central} object (BCG+ICL) within $\Rtwo$.
&
$f_{\rm BCG} \equiv \frac{M_{\rm \ast, BCG}}{M_{\rm \ast, BCG+ICL}} = \frac{M_{\rm \ast, BCG}}{M_{\rm \ast,cen}}$\\ 

\hline
\end{tabular}
\caption{Definitions of galaxy cluster, their stellar components and the mass fractions used throughout this paper. Measurements are for snapshot $z=0$. Note that depending on the assumed aperture radius, $r_{\rm ap}$, we actually use many definitions of BCG and ICL masses and fractions.}
\label{tab: Definition components}
\end{threeparttable}
\end{center}
\end{table*} 

\subsection{Definitions of Stellar Components and Fractions}
\label{subsec:stellar_components}

In Table \ref{tab: Definition components} we present the definitions of the different stellar components of the simulated clusters for which we measure their masses and fractions. All definitions are for 3D spherical distributions. In this work we are interested in studying the BCG+ICL or central object (the one associated with the main subhalo as determined by the \textsc{\large subfind} algorithm, that is, excluding satellite galaxies) and its separation into the BCG and ICL components.\footnote{Note that the nomenclature of BCG and ICL and its separation are nominal here since we are not measuring quantities related to light but to stellar mass, and we are not defining in a dynamical or evolutionary way the diffuse stellar particles associated with the ICL.}.
In Section~\ref{sec:radial-distribution} we explore several aperture radii, $r_{\rm ap}$, which were used in the literature to nominally separate the BCG and ICL components. 

Our main goal in this paper is to explore how the mass fractions of BCG+ICL and the BCG/ICL components in their different definitions depend on various cluster properties at $z=0$. At this point, it is important to clearly establish which fractions we are referring to. In theoretical and observational literature, the same term is used indiscriminately for different definitions. As shown in Table \ref{tab: Definition components}, it is particularly important to differentiate between ICL {\it total mass} fraction and ICL mass fraction, $F_{\rm ICL}$ and $\ficl$, respectively. In the former case the fraction is with respect to the total cluster stellar mass (BCG + ICL + satellites, that is, $\Mtwost$), while in the latter case it is rather with respect to the BCG + ICL mass; hence, $\ficl>F_{\rm ICL}$.

\subsection{Dynamical state of the clusters}
\label{sec:dynamical-stage}

Clusters tend to exist in varied stages of dynamical evolution. For our cluster sample, we define the dynamical state according to a prescription detailed in \citep[see also \citealp{zhang2022,marini2022ML}]{neto2007}, 
which is based on the center-of-mass offset ($\Delta_r$) and the subhalo mass fraction ($f_s$), where this fraction excludes the main subhalo.
These quantities are defined as follows:

\begin{itemize}
\item \eqmakebox[things][l]{Centre of Mass Offset: }
     $ \begin{aligned}[t]
      \Delta_r = \frac{\left|{\bf r}_{\rm min} -{\bf r}_{\rm cm}\right|}{R_{200}}
      \end{aligned} $
\item \eqmakebox[things][l]{Subhalo Mass Fraction: }
     $ \begin{aligned}[t]
      f_s = \frac{M_{\rm tot,sub}}{M_{200}},
      \end{aligned} $
\end{itemize}
where ${\bf r}_{\rm min} - {\bf r}_{\rm cm}$ represents the distance between the centre of mass ${\bf r}_{\rm cm}$\footnote{${\bf r}_{\rm cm}$ is the sum of the mass-weighted relative coordinates of all particles/cells within the main subhalo.} and the position of the minimum value of the gravitational potential (${\bf r}_{\rm min}$) , and $M_{\rm tot, sub}$ is the total mass in subhaloes, including stellar and gas cells. Adopting the threshold parameters proposed by \citet{biffi2016}, we classify the dynamical state of our clusters at $z=0$ into three cases: \textit{(i)} relaxed, when $\Delta_r < 0.07$ and $f_s < 0.1$; \textit{(ii)} intermediate, when only one of $\Delta_r < 0.07$ or $f_s < 0.1$ is satisfied; and \textit{(iii)} disturbed, when none of the previous criteria are satisfied (i.e., $\Delta_r > 0.07$ and $f_s > 0.1$). As a result, for our TNG300 sample, $\approx 16.3\%$ (113 clusters) are classified as relaxed, $\approx 28.4\%$ (199 clusters) as intermediate, and $\approx 55.4\%$ (388 clusters) as disturbed. 
The dynamical state of the cluster is expected to depend on the evolutionary stage of its DM halo, quantified, for example, by the epoch at which 50\% of the cluster's current mass was formed, $\zfive$. In the hierarchical clustering scenario, the more massive the structure, the later it is assembled on average. On the other hand, it is well known that the concentration of the halo depends on its assembly history, so that the higher the $\zfive$, the higher the concentration \citep[e.g.,][]{avila1999,wechsler2002}. This is mainly because concentration is related to the critical density of the universe at the time when the mass within the scale radius $r_s$ was assembled, a time associated with $\zfive$ \citep[e.g.,][]{ludlow2013}. Additionally, it captures the increase in concentration due to the reduction in scale radius over time as satellites merge with the central \citep[e.g. ][]{Ragagnin2019}. 
As usual, we define the halo concentration as $c_{200}\equiv\Rtwo/r_s$, where $r_s$ is the scale radius obtained by fitting the 3D halo density distribution\footnote{The measurements of $c_{200}$ (as well as of $\zfive$) that we use here are for the hydrodynamics simulations, i.e., they include baryons and their effects on the mass distribution. Moreover, the concentrations we use (taken from \citealp{anbajagane2022}), were measured for the central object/main subhalo. These authors show how the scale radii and concentrations in the IllustrisTNG hydrodynamical simulations compare with those in the DM-only ones. In the cluster mass range, the former have slightly lower concentrations than the latter with no dependence on $\Mtwo$ in this mass range.} to the NFW profile \citep{navarro1997NFW}.

\begin{figure}
  \centering
    \includegraphics[width=1.09\linewidth]{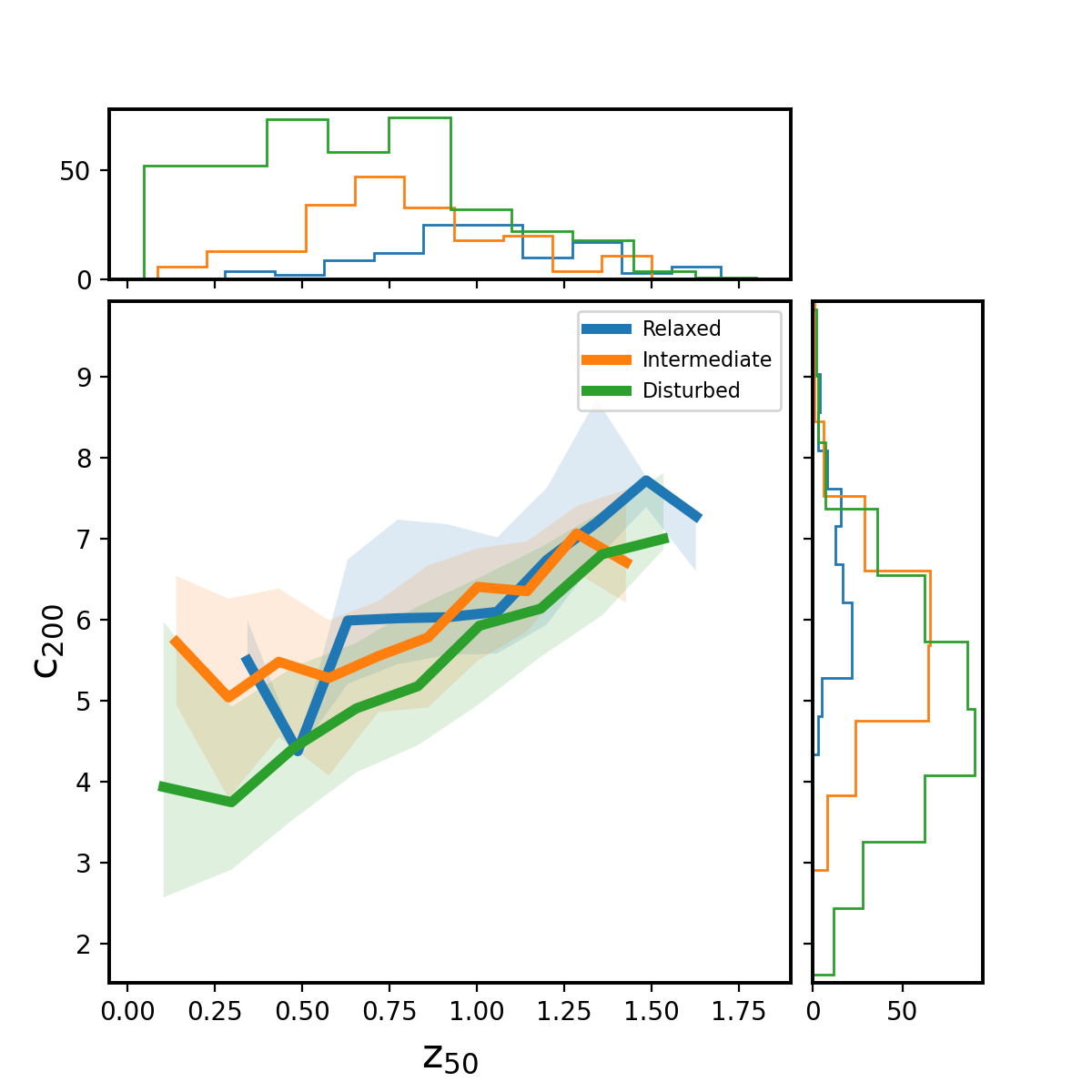}
	\caption{Halo concentration $\ctwo$ (see footnote 3) as function of halo assembly time $\zfive$ (the redshift at which the host halo had assembled 50 percent of its present-day mass), segregated by dynamical state: relaxed (blue), intermediate (orange), and disturbed (green) clusters. The solid lines are the median values and the shaded regions indicate the 16-84th percentile ranges.
    The upper and right plots present the corresponding histograms for $\zfive$ and $\ctwo$, respectively.}
	\label{fig:c200_vs_z50_dyn_state}
\end{figure}

Figure ~\ref{fig:c200_vs_z50_dyn_state} shows the aforementioned correlation of $c_{200}$ with $\zfive$ for our 700 TNG300 clusters separated into the three dynamical states. The green, orange, and blue lines correspond to the medians for the disturbed, intermediate, and relaxed clusters respectively, while the shaded areas correspond to the 16th and 84th percentile ranges, respectively. The corresponding distributions of $\zfive$ and $c_{200}$ are shown in the top and right panels, respectively. As expected, the more dynamically disturbed clusters assemble later and into less concentrated haloes. As mentioned above, our sample is dominated by disturbed clusters, which are actually in their active assembly phase. On the other hand, the $c_{200}$--$\zfive$ correlation is slightly segregated by the dynamical state: the more disturbed the cluster, the lower the $c_{200}$ for a given $\zfive$ on average. It is well known that measurement of $c_{200}$ in DM haloes after major mergers tends to give values lower than the haloes will have after they relax. However, concentration can also increase during the short-term pericentric passages of the merged substructure \citep{ludlow2012,klypin2016}, which introduces scatter in the $\zfive$--$c_{200}$ correlation. This is expected when the concentration is calculated for the entire halo, but can also occur when the concentration is based only on the main/central subhalo (as in our case) if the satellite subhalo liberates a substantial amount of its matter near the pericentre, or if \textsc{\large subfind} artificially assigns some of
the satellite's mass to the main subhalo or links the two when they are in close proximity \citep[e.g.][]{Knebe2011,Muldrew2011,Han2012,Behroozi2015}. Estimating the contribution from spurious effects
related to substructure discernment to the scatter in the $c_{200}$--$\zfive$ relation would require
an in-depth exploration with alternative halo-finding algorithms that may potentially alleviate
the drawbacks of \textsc{\large subfind}. This is beyond the scope of this paper, and
hence, we refrain from commenting further on this aspect.

\begin{figure}
  \centering
    \includegraphics[width=0.95\linewidth]{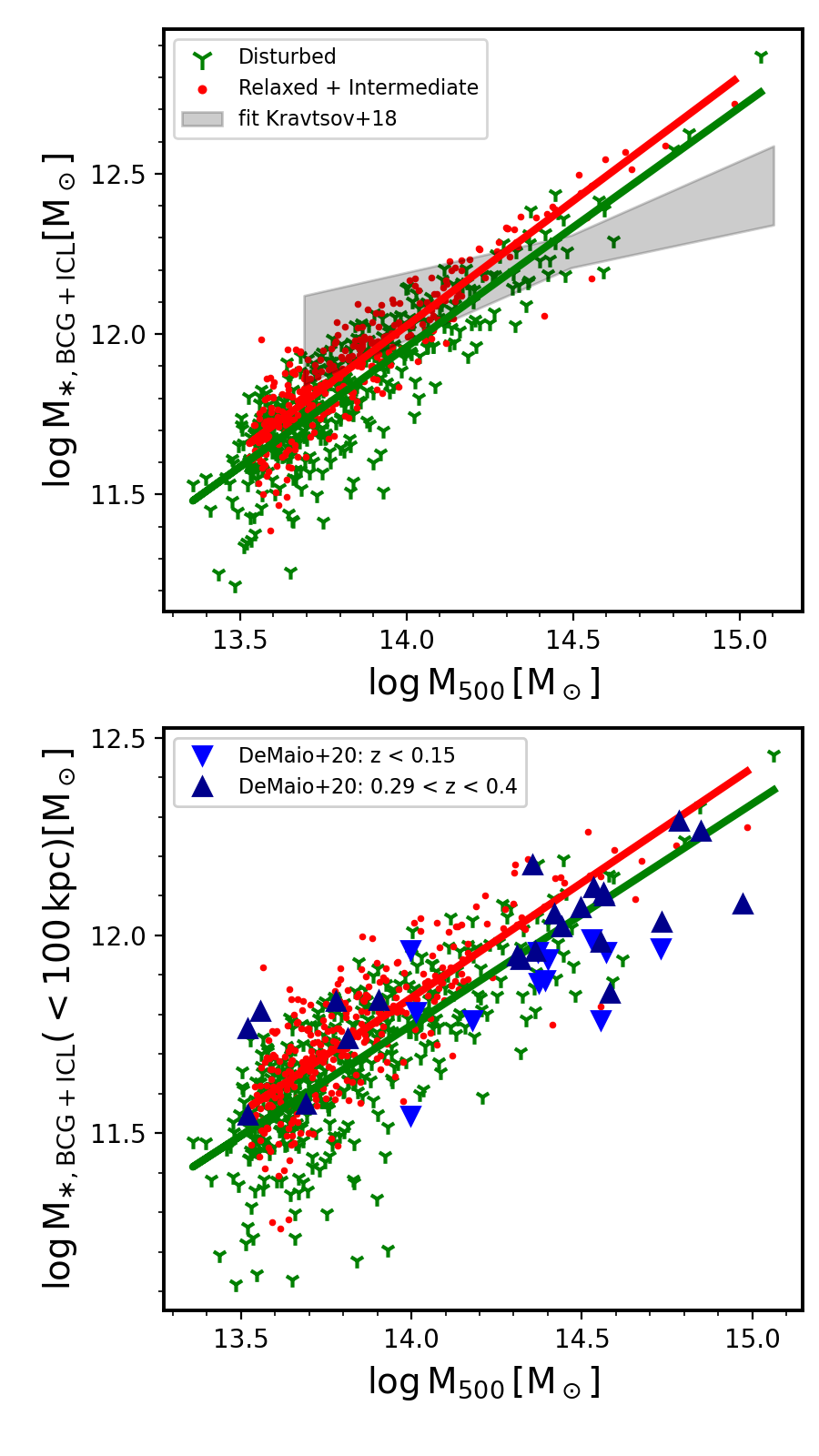}
	\caption{Stellar mass of the BCG+ICL or central object within the cluster radius ($\Rtwo$; top panel) and 100 kpc (bottom panel) as a function of cluster mass up to the overdensity $\Delta=500$, $\Mfive$. Clusters defined as perturbed (green tri down) and intermediate + relaxed (red dots) are shown separately. The shaded region in the top panel shows a linear fit and its dispersion to observations according to \citet{Kravtsov2018}. The upward (downward) triangles in the bottom panel correspond to estimates made for clusters observed at redshifts $z < 0.14 $ ($0.29 < z < 0.4$)
 }
	\label{fig:Mstar_bcg+icl_vs_logm500_comparison}
\end{figure}

\section{Radial stellar mass distributions of the BCG, ICL and satellite galaxies}
\label{sec:results}

In this Section, we present our main results from the analysis of the 3D stellar mass distribution at $z=0$ of the 700 TNG300 clusters of galaxies described in Section~\ref{sec:methodology}. We begin by comparing with observational determinations from the literature the BCG+ICL stellar masses, $M_{\rm \ast, BCG+ICL}$, as a function of the cluster mass $M_{500}$, and explore how the BCG+ICL mass fraction, $F_{\rm BCG+ICL}$, depends on several cluster properties (Section~\ref{sec:BCG+ICL}). In Section~\ref{sec:radial-distribution} we study the stellar mass distributions of the cluster central objects and explore the dependence of different determinations of the ICL mass fractions ($\ficl$), using different criteria for the aperture radii, on the cluster mass. 
Then, we explore in detail which cluster properties most affect both fractions, $\ficl$ and $F_{\rm ICL}$, by choosing a representative criterion for the aperture radius.

\subsection{Stellar masses and fractions of the BCG+ICL component}
\label{sec:BCG+ICL}

In Fig.~\ref{fig:Mstar_bcg+icl_vs_logm500_comparison} we present the stellar mass of the central objects, $M_{\rm \ast, BCG+ICL}$, versus the total cluster mass as obtained for our TNG300 cluster sample \cite[see also][\citetalias{montenegro2023}]{Pillepich2018a}. To compare these results with observational determinations, we use the cluster mass defined at $\Delta=500$ instead of 200, $M_{\rm 500}$. In both panels, we separate our sample into two dynamical states: disturbed (green tri down) and relaxed/intermediate (orange dots) clusters.
A linear regression to the data shows that $\log(M_{\rm \ast, BCG+ICL}/\msun)= a +b\log M_{\rm 500}$, with $a=1.50$ and $b=0.75$ for the disturbed clusters (green line) and $a=1.14$ and $b=0.78$ for the relaxed/intermediate ones (orange line). 
In the top panel, the fit and standard deviation reported in \citet{Kravtsov2018} for a sample of 20 observed clusters (including those reported in \citealp{gonzalez2013galaxy}) are plotted with the gray area. In the lower panel, we use the stellar mass distribution within 100 kpc for comparison with the observational determinations from \citet[][]{DeMaio2020}. 

According to Fig.~\ref{fig:Mstar_bcg+icl_vs_logm500_comparison}, the simulation results and observations are roughly consistent. Note that the comparison is not enterely fair: the former BCG+ICL masses are determined for 3D stellar mass distributions of the central object, while the latter are based from the observed (projected) surface brightnesses after excluding (masking) satellite galaxies. In more detail, the predicted correlation is steeper than the one from the observation.  Note that there are few clusters in the high mass range in both observed and simulated samples, so that the discrepancy in this mass range may be due to low statistics. On the other hand, observations of larger clusters might give lower luminosities (stellar masses) because they might not reach the outermost low surface brightness regions. Furthermore, by nature most massive clusters tend to be more disturbed because they assemble later. As our results show, relaxed/intermediate clusters have higher BCG+ICL stellar masses on average than disturbed ones. The very few observed massive clusters (likely disturbed) actually have a BCG+ICL stellar mass more in line with disturbed TNG300 clusters than with relaxed/intermediate ones.

\begin{figure}
  \centering
    \includegraphics[width=0.95\linewidth]{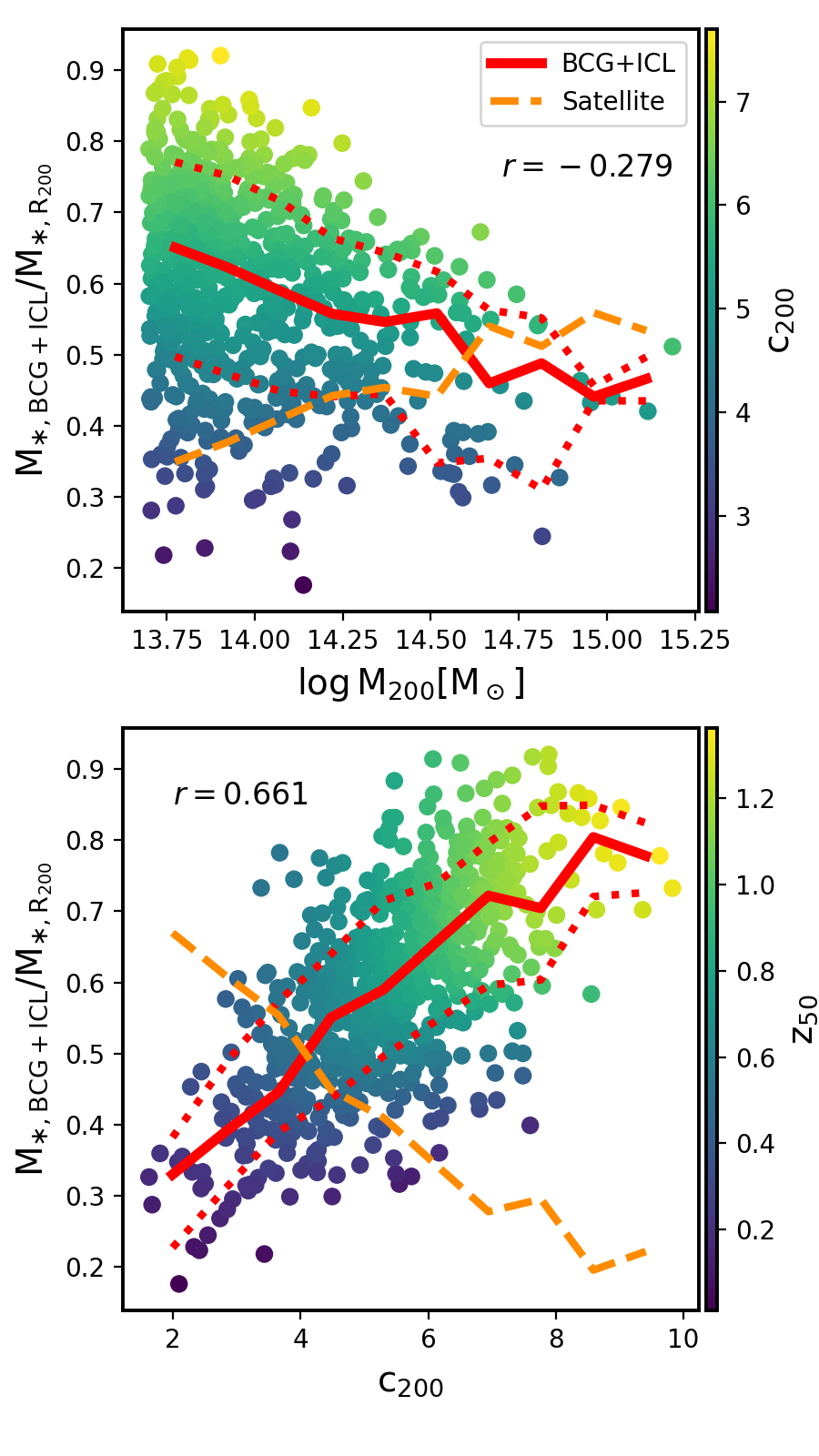}
	\caption{Stellar mass fraction of the BCG+ICL or central object as a function of $\Mtwo$ (top panel) and NFW concentration $\ctwo$ (bottom panel). The colour scale in the top and bottom panels represents $\ctwo$ and half-mass assembly redshift $z_{50}$, respectively. The colour of the data points is smoothed using the LOESS algorithm. The corresponding median and 16-84th percentiles are shown with red solid and dotted lines, respectively. The corresponding Pearson correlation coefficients $r$ are indicated inside the panels.  The dashed orange lines are the medians of the complement to $F_{\rm BCG+ICL}$, i.e. the satellite stellar mass fraction.}
	\label{fig:scatter_f_bcg_icl_vs_m200-c200_cc200-cz50}
\end{figure}

Before discussing how to separate the ICL from the BCG, we explore how the BCG+ICL mass fraction, $F_{\rm BCG+ICL}$, correlates with various cluster properties.
Since BCG and ICL share stellar particles spatially, making it difficult to separate them with a simple radial cut, some authors have suggested treating the two components as one, rather than treating them separately \citep{gonzalez2007,Kravtsov2018,zhang2019,Chun+2023}. 
The top panel of Fig.~\ref{fig:scatter_f_bcg_icl_vs_m200-c200_cc200-cz50} shows $F_{\rm BCG+ICL}$ as a function of $\Mtwo$ for our TNG300 cluster sample at $z=0$. The median and 16th-84th percentiles are plotted with the red solid and dotted lines, respectively. The dashed orange line shows the median of the complement of $F_{\rm BCG+ICL}$, $F_{\rm sat}= 1 -F_{\rm BCG+ICL}$, i.e., the mass fraction of the satellites. A noisy trend of decreasing BCG+ICL (increasing satellite) mass fractions is seen with increasing $\Mtwo$, consistent with observational determinations \citep{gonzalez2007,gonzalez2013galaxy, Kravtsov2018}. There are two factors at play in this trend: (i) massive haloes have assembled more recently, and
hence, have had less time available for ICL formation, and (ii) massive haloes have lower concentrations on average, implying a weaker tidal field and dynamical friction experienced by satellites.

The median $F_{\rm BCG+ICL}$ decreases from $\sim 0.65$ to $\sim 0.45$ in the mass range from $\Mtwo\approx 6\times 10^{13}$ $\Msun$ to $\Mtwo\approx 10^{15}$ $\Msun$. However, the (anti)correlation exhibits considerable scatter, with a Pearson correlation coefficient $r=-0.279$. The ICL and a significant fraction of the BCG are believed to be composed mostly of ex situ stellar particles \citep[e.g.,][\citetalias{montenegro2023}]{Pillepich2018a}, which come from the tidal striping and disruption of the cluster's satellite galaxies. The tidal field and dynamical friction generated by the cluster's potential is stronger for higher concentrations. Therefore, we expect that for a given cluster mass, $F_{\rm BCG+ICL}$ will correlate with the cluster concentration, $c_{200}$. In the top panel of Fig.~\ref{fig:scatter_f_bcg_icl_vs_m200-c200_cc200-cz50} we colour code the data by $c_{200}$. For illustrative purposes, we emphasise the presence/absence of residual trends by smoothing the colour distributions using the locally weighted regression algorithm LOESS \citep{cleveland1988, cappellari2013}. The segregation of $F_{\rm BCG+ICL}$ in this plane is strong, and in the sense that for given $\Mtwo$, the higher the $c_{200}$, the larger the BCG+ICL mass fraction. On the other hand, we do not observe a correlation between $c_{200}$ and $\Mtwo$ in our mass range (see also footnote 4).

In the bottom panel of Fig.~\ref{fig:scatter_f_bcg_icl_vs_m200-c200_cc200-cz50} we plot now $F_{\rm BCG+ICL}$ as a function of $c_{200}$. The correlation ($r=0.661$) is clearly stronger than with $\Mtwo$, varying the median of $F_{\rm BCG+ICL}$ (red solid line) from $\sim 0.35$ to $\sim 0.8$ in the concentration range from $c_{200}\approx 2$ to $\approx 9$. In contrast, the mass fraction in satellites decrease strongly with $c_{200}$ (see orange dashed line, which shows the median of this fraction as a function of $c_{200}$). The data is colour coded by $z_{50}$ using the smoothing procedure of the LOESS algorithm. 
In agreement with Fig.~\ref{fig:c200_vs_z50_dyn_state},  $c_{200}$ correlates with $z_{50}$, but for a given $c_{200}$, the higher the $z_{50}$, the larger the $F_{\rm BCG+ICL}$. The above imply that for a similar cluster gravitational tidal field (approximately similar $c_{200}$), satellite
galaxies experienced tidal stripping for longer duration within the clusters that assembled earlier, reducing the mass budget in satellites in favour of mass growth for both the BCG and the ICL.

\begin{figure}
  \centering
    \includegraphics[width=1.\linewidth]{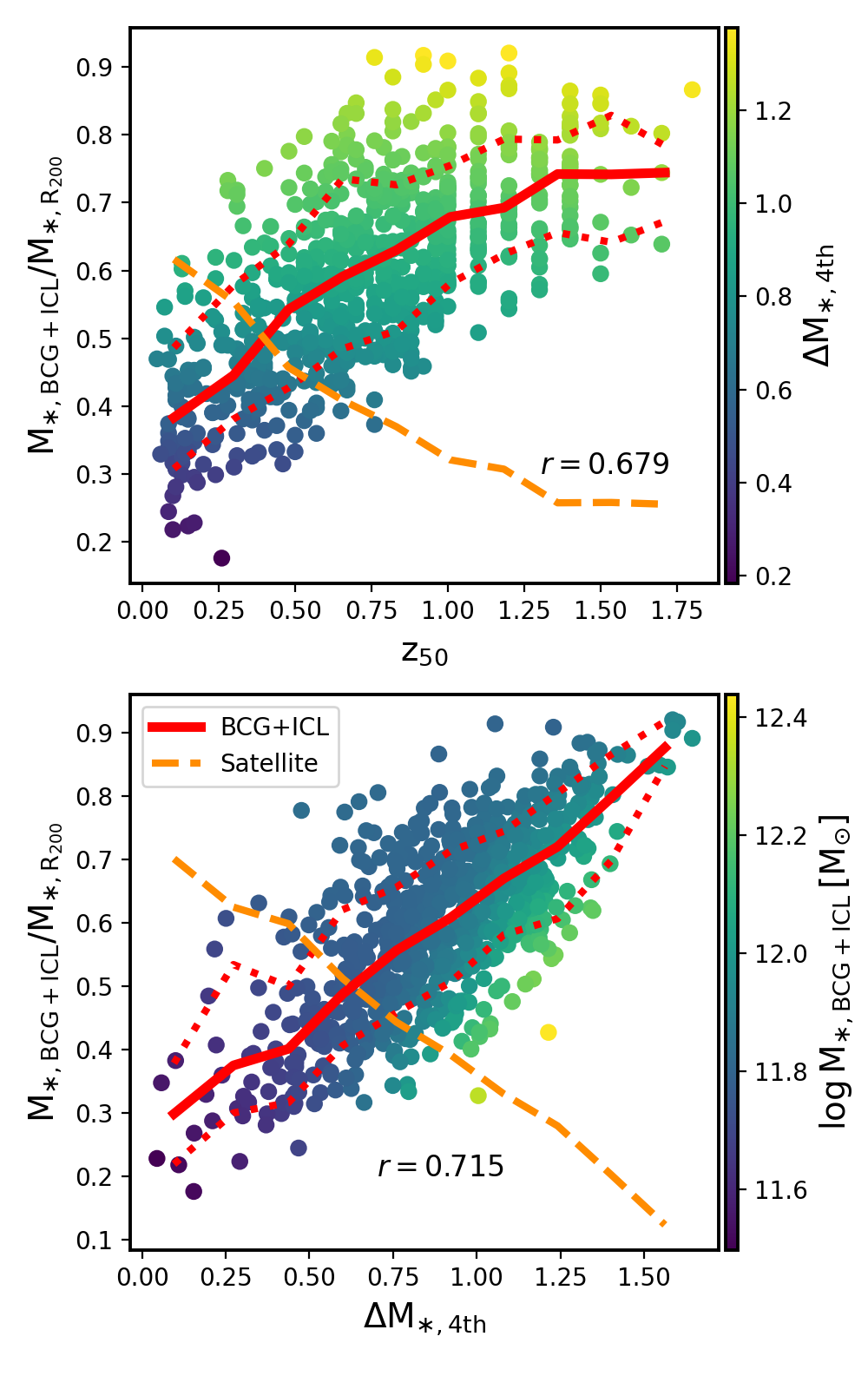}
	\caption{Stellar mass fraction of the BCG+ICL or central object as a function of $z_{50}$ (top panel) and mass gap with the fourth more massive satellite, $\deltam$ (bottom panel). The colour scale in the top and bottom panels represents $\deltam$ and BCG+ICL or central object mass, respectively. The colour of the data points is smoothed using the LOESS algorithm. The corresponding median and 16-84th percentiles are shown with red solid and dotted lines, respectively. The corresponding Pearson correlation coefficients $r$ are indicated inside the panels. The dashed orange lines are the medians of the complement to $F_{\rm BCG+ICL}$, i.e. the satellite stellar mass fraction. 
}
	\label{fig:scatter_f_bcg_icl_vs_z50-deltam4_cdeltam4-cz50}
\end{figure}

In the top panel of Fig.~\ref{fig:scatter_f_bcg_icl_vs_z50-deltam4_cdeltam4-cz50}, $F_{\rm BCG+ICL}$ is now plotted as a function of $z_{50}$. As mentioned above, there is a clear correlation between these two quantities, with larger values of $F_{\rm BCG+ICL}$ indicating earlier assembly \citep[higher $z_{50}$; see for a similar result,][]{Chun+2023}. The Pearson correlation coefficient ($r=0.679$) is very similar to that of $c_{200}$, which implies that $c_{200}$ and $z_{50}$ correlate, but also that the growth of BCG + ICL mass is equally influenced by the cluster tidal field and its history of assembly, which are separately traced by $c_{200}$ and $z_{50}$. Also, since earlier assembled clusters tend to be more relaxed at $z=0$ (see Fig.~\ref{fig:c200_vs_z50_dyn_state}), we see a trend towards higher BCG+ICL stellar mass fractions in more relaxed clusters. We searched for an observable quantity that correlates best with $F_{\rm BCG+ICL}$, and found it to be the mass gap between the most massive and the fourth most massive member galaxy within $\Rtwo$: $\deltam=\log(M_{1}/M_{4})$. 
In the top panel of Fig.~\ref{fig:scatter_f_bcg_icl_vs_z50-deltam4_cdeltam4-cz50} the data is colour coded by $\deltam$\footnote{Similar results are obtained using the second or third most massive member, but the segregation is stronger for the fourth. For higher masses, the differences are minimal. On the other hand, the luminosity gap using the fourth or fifth member has been found to provide a more robust correction for predicting the cluster mass given the BCG luminosity \citep{lu2016}.}. 
For a given $z_{50}$, the larger $\deltam$, the higher the $F_{\rm BCG+ICL}$. On the other hand, there is a trend of increasing $\deltam$ as the cluster assembles earlier, that is, when $z_{50}$ is higher. This is to be expected, as the earlier the cluster is assembled, the more the BCG can grow through mergers with massive cluster members, widening the gap.

In the bottom panel of Fig.~\ref{fig:scatter_f_bcg_icl_vs_z50-deltam4_cdeltam4-cz50}, the correlation between $F_{\rm BCG+ICL}$ and $\deltam$ is plotted. The correlation is tighter than any of the previous ones ($r=0.715$): the median of $F_{\rm BCG+ICL}$ increases from $\approx0.3$ to $\approx 0.85$ for the lowest to highest gap values. 
We colour code the data by $M_{\rm \ast, BCG+ICL}$, a quantity that eventually can be determined from observations. The above shows that the scatter of the $F_{\rm BCG+ICL}$--$\deltam$ correlation segregates by $M_{\rm \ast, BCG+ICL}$: for a given $\deltam$ gap, the larger $M_{\rm \ast, BCG+ICL}$, the lower $F_{\rm BCG+ICL}$. This is in part due to the strong correlation between $M_{\rm \ast, BCG+ICL}$ and $\Mfive$, and the fact that massive clusters have smaller $F_{\rm BCG+ICL}$ values.

Taken together, we conclude that the BCG+ICL mass fraction is mainly controlled by the halo concentration, which is in turn set mainly by its assembly history (partially quantified by its formation time, $z_{50}$). Furthermore, the concentration correlates well with the mass gap using the 4th most massive member, $\deltam$, with the scatter modulated by the BCG+ICL mass – both of which can eventually be inferred from deep photometric observations.

\begin{figure*}
  \centering
  \vbox{
    \includegraphics[width=0.95\linewidth]{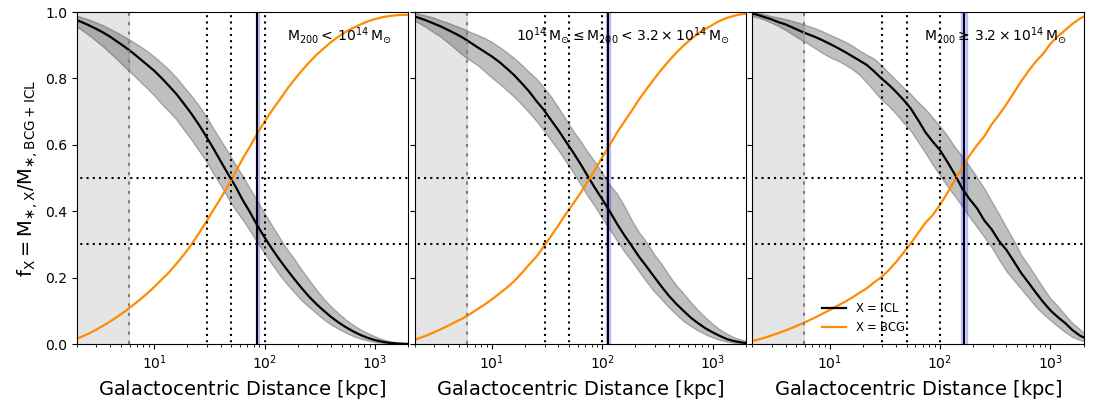}
	\\
    \includegraphics[width=0.95\linewidth]{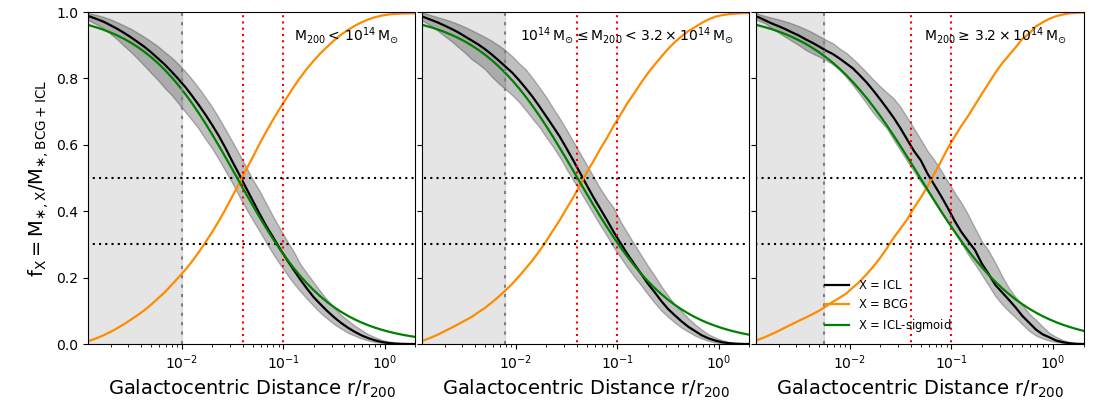}
    \\
    \includegraphics[width=0.95\linewidth]{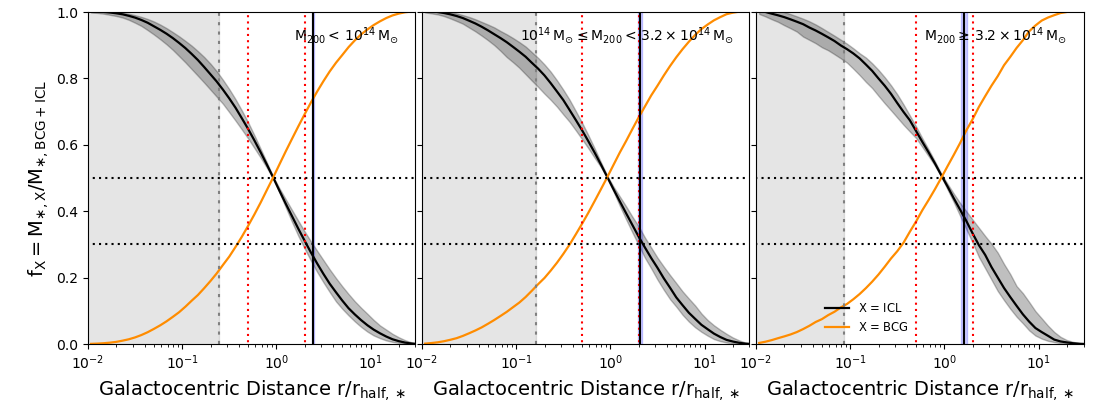}
    }
    \caption{Median normalised stellar mass profiles (or BCG/ICL mass fractions) of the BCG, orange solid line, and ICL, black solid line, at $z = 0$ as a function of the used aperture radius (from very inner to very outer clustercentric radii), for different cluster mass ranges: $\Mtwo = 0.5$--$1 \times 10^{14} \, \Msun$ (left), $\Mtwo = 1$--$3.2 \times 10^{14}\, \Msun$ (medium) and $\Mtwo \geq 3.2 \times 10^{14} \, \Msun$ (right). The plots can be read as ``tell me the aperture radius and I tell you the median mass fraction of BCG or ICL at that radius'', $f_{\rm BCG}$ and $\ficl$, respectively. In the upper, medium and lower panels the aperture radius is in units of kpc, $\Rtwo$, and \rhalf, respectively. The shaded regions around the black solid line show the 16th-84th percentile ranges in the case of the ICL component. The vertically grey shaded region indicates the resolution limit. In the upper panels, the vertical dotted lines indicate 30, 50, and 100 kpc, and the solid black lines indicate the mean 0.1$\Rtwo$ radius for each mass bin, with the narrow blue shaded region showing the standard deviation. In the medium panels, the red dotted lines indicate 0.05 and 0.1 $\Rtwo$. The green lines are the complement to the BCG+ICL analytic mass profile given in \citep[][see text]{Pillepich2018a}. In the lower panels, the red dotted lines indicate  0.5 and 2 \rhalf; the black solid lines and narrow blue shaded areas around them are as in the upper panels.}
	\label{fig:median_profile_f_bcg_icl_kpc}
\end{figure*}

\subsection{Radial stellar mass distributions and BCG/ICL separation}
\label{sec:radial-distribution}

As mentioned in Sec. \ref{sec: Intro}, the separation into BCG and ICL components is a non-trivial task from the observational point of view.
From a theoretical point of view, several studies \citep[e.g.,][\citetalias{montenegro2023}]{contini2014formation,Pillepich2018a} have shown that the stellar particles associated with both BCG and ICL are well mixed in a region that can be quite extended, making it unfeasible to determine a clear breakup radius that separates the BCG from the ICL. 
However, it is desirable to try to find a \textit{nominal} transition radius between these two components. 
In this spirit, fixed physical radii for all clusters (e.g., 30, 50 or 100 kpc) or radii that scale with typical cluster sizes have been used in the literature. Examples of the latter case are a fraction of the cluster halo radius or a radius in units of the half-light (or half-mass) radius, \rhalf, of the BCG+ICL object. Following these definitions, here we will explore different aperture radii in the 3D spherically-averaged stellar mass distribution, up to which and beyond which we define the masses of the BCG and ICL components, respectively (see Table \ref{tab: Definition components}).

We begin by analyzing the spherically-averaged stellar mass profile of the central object. For any given transition radius, $r_{\rm ap}$, we associate the stellar mass located within the corresponding aperture to the BCG, and the mass beyond the aperture, but within $R_{200}$, to the ICL. Normalizing these masses by the central object mass, $M_{\rm \ast,BCG+ICL}$ within $\Rtwo$, gives the BCG and ICL mass fractions, $f_{\rm BCG}$ and $f_{\rm ICL}=1-f_{\rm BCG}$, respectively (see Table \ref{tab: Definition components}).

\begin{figure*}
  \centering
    \includegraphics[width=\linewidth]{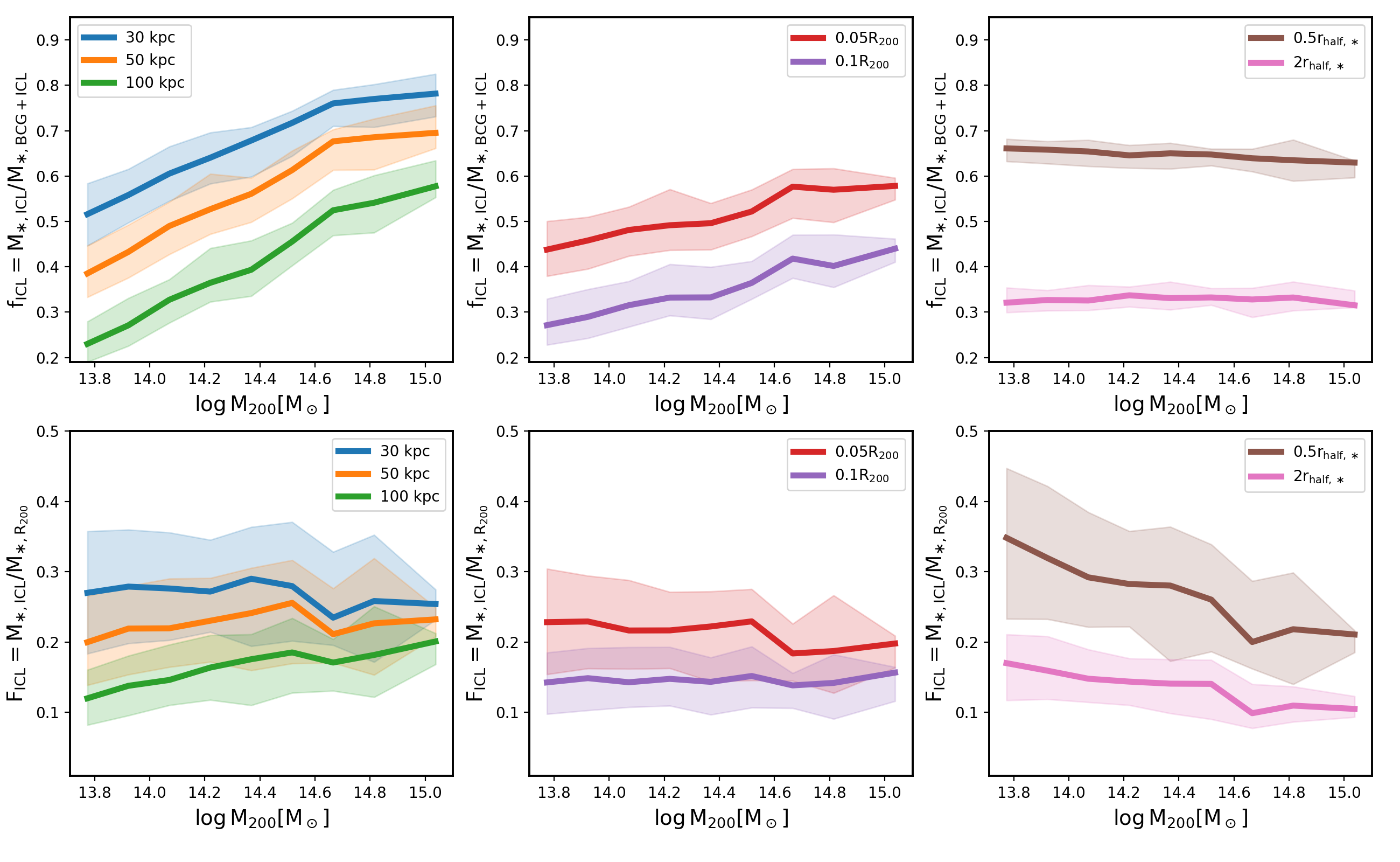}
	\caption{Median and 16-84th percentile range of ICL stellar mass fractions (upper panels) and ICL total stellar mass fractions (lower panels) as a function of $\Mtwo$ (see Table \ref{tab: Definition components} for the definitions). In the left panels, the aperture radius for separating the ICL from the BCG is defined at 30, 50 and 100 kpc. In the medium panels, this radius is defined at 0.05 and 0.1 $\Rtwo$, and in the right panels is defined at 0.5 and 2 \rhalf. Note how the ICL fraction values and their dependence on $\Mtwo$ change depending on the aperture radius used to separate the ICL from the BCG, and depending on the definition of the ICL fraction, either $\ficl$ or $F_{\rm ICL}$ (see Table \ref{tab: Definition components}.}
 
\label{fig:median_f_F_icl_vs_m200_diff_appertures}
\end{figure*}

In Fig.~\ref{fig:median_profile_f_bcg_icl_kpc} we present the median and 16th-84th percentile range of the ICL mass fraction, $f_{\rm ICL}$, as a function of a given aperture radius (from very inner to very outer clustercentric radii) in three halo mass bins: $\Mtwo = 0.5-1 \times 10^{14} \, \Msun$ (left), $\Mtwo = 1-3.2 \times 10^{14}\, \Msun$ (centre) and $\Mtwo = 3.2-15.3 \times 10^{14} \, \Msun$ (right). In the upper panels, the radius is given in kpc, in the middle panels it is given in units of $\Rtwo$, while in the lower panels it is given in units of the half-mass radius, \rhalf. For completeness, the orange curves show the respective medians of the BCG mass fractions, $f_{\rm BCG}$, i.e., the complement to $\ficl$.
The vertical grey shaded areas indicate the convergence radius of the simulation, which here we define as four times the gravitational softening length in TNG300 at $z=0$, $4\epsilon \approx 5.9 \, {\rm kpc}$ (see \citealt{Pillepich2018, Pillepich2018a} for additional comments on the effects of numerical resolution on galaxies' stellar masses and their spatial distribution). 
The vertical dotted black lines in the upper panels indicate radii commonly used in the literature for separating the BCG from the ICL, 30, 50, and 100 kpc, while the vertical solid black lines show the mean 0.1$\Rtwo$ of the clusters in the corresponding mass bin with the blue shaded region indicating their standard deviations.  
The two vertical dotted red lines in the intermediate panels indicate the radii of $0.05\Rtwo$ and $0.1\Rtwo$, while in the lower panels indicate the radii of 0.5 and 2\rhalf. 

The orange curves in Fig.~\ref{fig:median_profile_f_bcg_icl_kpc} are actually the central object cumulative mass profiles, which were also presented in \citet{Pillepich2018}. These authors showed that they can be well-fitted by a sigmoid function whose parameters depend on $\Mtwo$.
As seen in the upper and medium panels of Fig.~\ref{fig:median_profile_f_bcg_icl_kpc}, these cumulative profiles in physical units (kpc) or in units of $\Rtwo$, become shallower as $\Mtwo$ is larger, in agreement with the fits given in \citet[][]{Pillepich2018}\footnote{The green line in the medium panels of Fig.~\ref{fig:median_profile_f_bcg_icl_kpc} show the complement to the fitted sigmoid function given in \citet{Pillepich2018}, that is, $1-M_{\rm \ast,cen}(<r/\Rtwo)/M_{\rm \ast,cen}(\Rtwo)$, where we used the mean $\log\Mtwo$ in each panel for the formula. This is in fact that we call the ICL mass fraction, $\ficl$, see Table \ref{tab: Definition components}. The fit is a relatively good approximation to the simulation results}; the stellar half-mass radius, where $\ficl=f_{\rm BCG}=0.5$ if used as the transition radius, increases with $\Mtwo$ slightly faster than $\Rtwo$ does, implying that the BCG+ICL mass profile becomes more extended (shallower) on average as $\Mtwo$ increases. 
However, as expected, in units of \rhalf, the cumulative mass profiles are independent of the cluster mass (lower panels), which implies that the ICL mass fraction, $\ficl$ (or its complement, $f_{\rm BCG}$), at a given aperture radius in units of \rhalf\ is independent of $\Mtwo$. The BCG/ICL cumulative mass profiles are almost self-similar when radially scaled to \rhalf. This self-similarity is somewhat by construction due to the way \rhalf\ is defined. 

As a consequence of the above, $\ficl$ for a fixed aperture radius in physical units (30, 50, 100 kpc) or at a given fraction of $\Rtwo$, increases with $\Mtwo$, while this fraction measured at a radius that scales with \rhalf, is nearly independent of $\Mtwo$. On the other hand, the radius both in kpc or in units of, e.g., $0.1\Rtwo$, where a fixed value of $\ficl$ is attained (e.g. 0.3 or 0.5) increases with $\Mtwo$, while in units of \rhalf, the fraction is nearly the same at all masses, to the point that at 1\rhalf, $\ficl=0.5=$const. by definition. 
In the next subsection, we study in detail the dependences of the ICL mass fractions $\ficl$ and $\ficltot$ defined with different aperture radii, on $\Mtwo$, as well as on what other cluster properties may depend these fractions.

\subsection{ICL stellar mass fractions}
\label{sec:ICL-fractions}

Figure \ref{fig:median_f_F_icl_vs_m200_diff_appertures} shows the measured ICL mass fraction, $\ficl$ (upper panels) and \textit{total} ICL mass fraction $\ficltot$ (lower panels), as a function of cluster mass, $\Mtwo$, for the different aperture radii mentioned above. The solid lines correspond to the median values of our cluster sample and the shaded regions indicate the 16th and 84th percentiles. 
The ICL mass fractions with respect to the BCG+ICL mass, $\ficl$, and to the total stellar mass, $\ficltot$ (see Table \ref{tab: Definition components}), are related as follows: 
\begin{equation}
    F_{\rm ICL}\equiv \frac{M_{\rm ICL}}{M_{\rm \ast,200}}=\frac{M_{\rm ICL}}{M_{\rm \ast,cen}}\times \frac{M_{\rm \ast,cen}}{M_{\rm \ast,200}}\equiv\ficl\times F_{\rm BCG+ICL}.
    \label{eq:F-f}
\end{equation}
Since $F_{\rm BCG+ICL}<1$, then $\ficltot<\ficl$. In Figs. \ref{fig:scatter_f_bcg_icl_vs_m200-c200_cc200-cz50} and \ref{fig:scatter_f_bcg_icl_vs_z50-deltam4_cdeltam4-cz50} we explored how $F_{\rm BCG+ICL}$ correlates with several cluster properties. In particular, we have found that $F_{\rm BCG+ICL}$ decreases with $\Mtwo$, though with a large scatter, while the strongest correlations are with $c_{200}$ (or $z_{50}$) and the gap $\deltam$.

According to Fig.~\ref{fig:median_f_F_icl_vs_m200_diff_appertures}, when defining a fixed aperture radius, $r_{\rm ap}$, in physical units (kpcs), $\ficl$ increases with $\Mtwo$, with an approximate dependence in our mass range of $\ficl\propto \Mtwo^{b}$, with $b=0.22,$ 0.26, and 0.29, for $r_{\rm ap}=30$, 50, and 100 kpc, respectively, and a normalization that increases as $r_{\rm ap}$ is smaller. As for $\ficltot$, according to equation~(\ref{eq:F-f}), since $\ficl$ increases with $\Mtwo$ while $F_{\rm BCG+ICL}$ decreases, $\ficltot$ should be nearly independent on $\Mtwo$ but with a large scatter. This is what we observe in the lower left panel of Fig.~\ref{fig:median_f_F_icl_vs_m200_diff_appertures}. As $r_{\rm ap}$ increases from 30 to 100 kpc, there is tendency for $\ficltot$ to increase slightly with mass and for the scatter to decrease. 

Since our sample of clusters covers a wide range of masses and radii, it may not be very appropriate to use a fixed aperture radius to separate the ICL from the BCG: while a given radius may contain the entire BCG for small systems, the same radius may be well inside the BCG for large systems. An aperture radius that scales with the size of the system may be more suitable; for example, a radius that is a fraction of the cluster's virial radius or a radius characteristic of the mass distribution of the central object (BCG+ICL). In the upper middle and right panels of Fig.~\ref{fig:median_f_F_icl_vs_m200_diff_appertures}, we show $\ficl$ versus $\Mtwo$ using $r_{\rm ap}=0.05-0.1\ \Rtwo$ and $r_{\rm ap}=0.5-2$  \rhalf, respectively. In the former case, $\ficl$ still increases with $\Mtwo$, but less than in the case of a fixed aperture radius, approximately $\ficl\propto \Mtwo^{0.17}$, while in the latter case, $\ficl$ is nearly independent of mass, in particular for $r_{\rm ap}\geq 1$ \rhalf. As discussed in the previous subsection, the BCG/ICL cumulative mass profiles scaled to \rhalf\ are self-similar. For $r_{\rm ap}=2$ \rhalf, $\ficl\approx 0.33$ with a very small scatter around this value (see Section~\ref{sec:ICL-fraction-dependancies} for its distribution) and, by definition, $\ficl=0.5=$const. when $r_{\rm ap} =$ \rhalf. 

Regarding the ICL total mass fraction, $\ficltot$, it is nearly independent of $\Mtwo$ for $r_{\rm ap}=0.05-0.1 \Rtwo$, but with large scatters. In the case of $r_{\rm ap}=0.5-2$ \rhalf, since $F_{\rm BCG+ICL}$ decreases with $\Mtwo$ (Fig.~\ref{fig:scatter_f_bcg_icl_vs_m200-c200_cc200-cz50}), $\ficltot$ should anti-correlate with $\Mtwo$ [see equation~(\ref{eq:F-f})]. This is what we observe in the lower right panel of Fig.~\ref{fig:median_f_F_icl_vs_m200_diff_appertures}.

In the literature, one or the other fraction ($\ficl$ or $\ficltot$) is often used interchangeably to quantify the amount of ICL in galaxy clusters. Figure \ref{fig:median_f_F_icl_vs_m200_diff_appertures} clearly shows how different the values of these fractions can be, regardless of the criterion used to separate the ICL from the BCG. According to equation~(\ref{eq:F-f}), $\ficltot<\ficl$ always. While it is important to differentiate between these two definitions of ICL mass fraction, the more important question now is which of the aperture radius-based criteria is more appropriate to nominally separate ICL from BCG. In the next section we address this issue.

\begin{figure}
  \centering
    \includegraphics[width=0.97\linewidth]{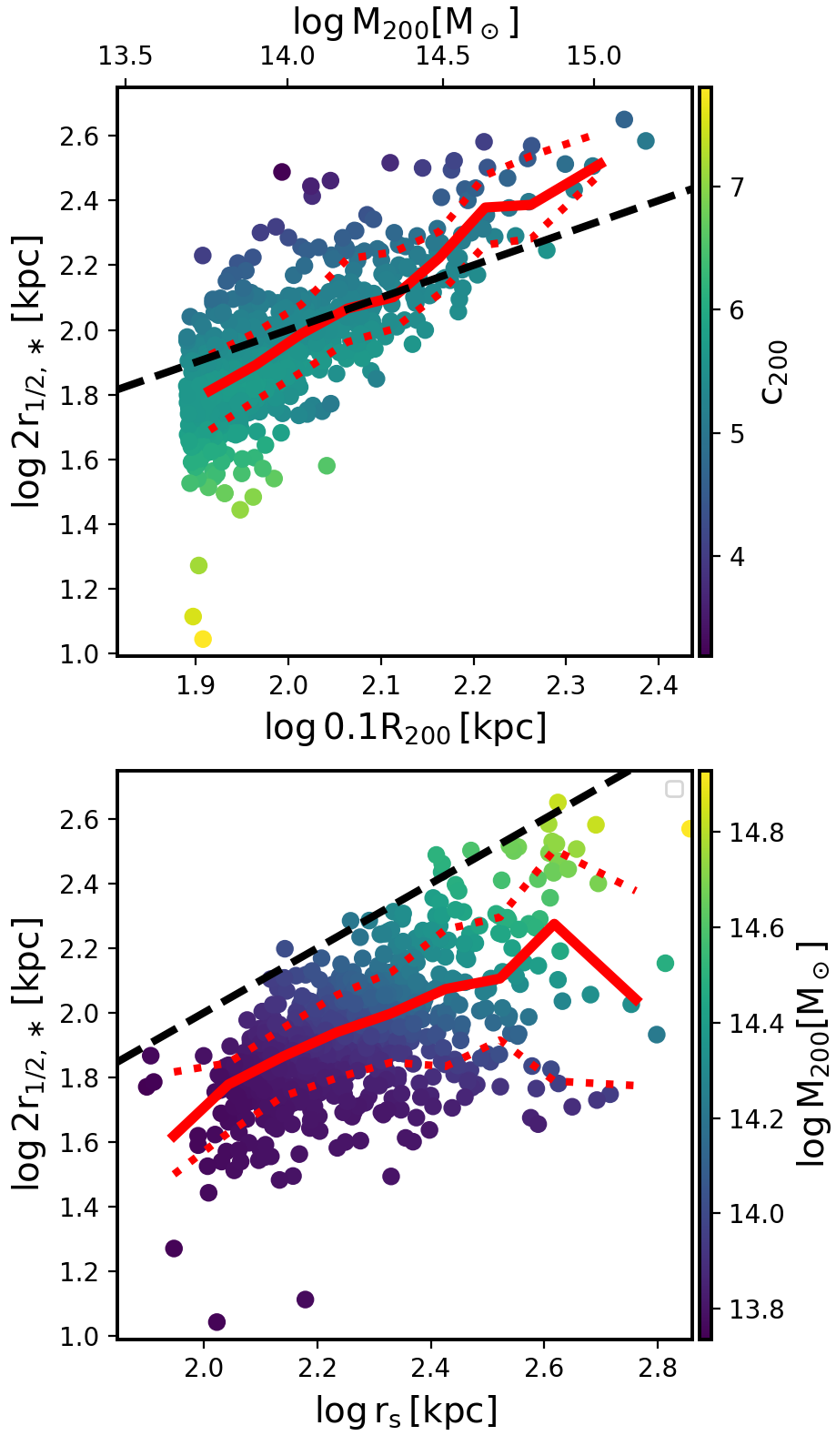}
	\caption{The stellar half-mass radius versus 0.1$\Rtwo$ (top panel, lower X axis),  $\Mtwo$ (top panel, upper X axis) and the NFW scale radius, $r_s$ (bottom panel), for our TNG300 cluster sample. The colour scale in the top and bottom panels represents $\ctwo$ and $\Mtwo$, respectively. The colour of the data points is smoothed using the LOESS algorithm. The corresponding medians and 16-84th percentiles are indicated by the red solid and dotted lines, respectively. The dashed line indicate the one-to-one relationship.}

	\label{fig:scatter_rhalf_m200_cc200_cm200}
\end{figure}

\subsection{Defining a transitional zone between the BCG and the ICL}
\label{sec:optimal-radius}

As discussed above, ICL and BCG stars overlap spatially in such a way that formally there is no breakup or aperture radius that separates both components. However, we can attempt to define a nominal transitional zone, characterized by a radius, where the ICL distribution overcomes the BCG’s one \citep[e.g.,][]{Chen+2022,contini2022,proctor2024}. Optimally, this radius would actually mark a zone where the gravitational influence of the BCG (baryons + dark matter) decreases substantially, that is, the stars are no longer bound to the central galaxy, and their velocity dispersion begins to be associated with the gravitational field of the entire cluster. 
Estimating this possible transitional radius is beyond the scope of this paper, but the radius 2\rhalf\ might be a natural and easily measured candidate, as will be discussed below.

The top panel of Figure \ref{fig:scatter_rhalf_m200_cc200_cm200} illustrates the relationship between 2\rhalf\ and $\Rtwo$ (lower X axis) and $\Mtwo$ (upper X axis). 
The solid red lines represent the median values, while the dotted lines correspond to the 16th and 84th percentiles. The data points are color-coded according to $c_{200}$.
The radius 2\rhalf\ scales with the size of the cluster (its total mass or virial radius). 
Roughly, 2\rhalf$\propto \Mtwo^{0.57}$ and 2\rhalf$\propto 0.1\Rtwo^{1.72}$ with a scatter that correlates with $c_{200}$. 
The fact that 2\rhalf\ increases with $0.1\Rtwo$ faster than a linear relation implies that the larger the cluster, the puffier (larger 2\rhalf) is the BCG+ICL central mass distribution (see also Fig.~\ref{fig:median_profile_f_bcg_icl_kpc}). On the other hand, for a given $\Mtwo$ or 0.1$\Rtwo$, the higher the $c_{200}$, the smaller the 2\rhalf. A higher cluster concentration implies stronger tidal forces and a deeper gravitational potential in the inner regions, so that the central stellar mass distribution becomes more concentrated (smaller 2\rhalf\ radius).

In the bottom panel of Fig.~\ref{fig:scatter_rhalf_m200_cc200_cm200} we show the correlation between 2\rhalf\ and the NFW scale radius, $r_s$, of the cluster dark matter component. The smoothed colour code indicates the cluster mass, $\Mtwo$. The escape velocity within $r_s$ is relatively high, while it decreases faster at larger radii. Therefore, in the vicinity of the BCG, within $r_s$, most of stars that are tidally disrupted/stripped from satellites after periapsis passage or ejected after BCG-satellite major mergers, cannot escape to larger radii. As discussed in \citet[][]{Chen+2022}, a fraction of them would fall back on to the BCG and increase its mass, while the rest would remain weakly bound to the BCG (forming probably its stellar halo) and form a transitional zone well confined within $r_s$, hence the BCG's sphere of influence.\footnote{After this paper was submitted, appeared a work by \citet{contini2024}, where, using a semi-analytical model, the authors calculate a halo mass-dependent radius encompassing the central galaxy and its stellar halo. This radius is taken by them as the stellar halo radius and defines the transition region towards the ICL component. Interesting enough, the transition radii they infer as a function of halo mass roughly agree with the 2\rhalf\ radii shown in Fig. \ref{fig:scatter_rhalf_m200_cc200_cm200} and with the 'intra-halo light' radii determined by \citet{proctor2024} from Hydrodynamical simulations (see below).}
The characteristic radius of this zone is primarily shaped by the growth of the BCG, but helped maintain by the inner potential of the dark matter halo, so that this radius should stay roughly within $r_s$  \citep[][]{Chen+2022}. 
According to Fig.~\ref{fig:scatter_rhalf_m200_cc200_cm200}, 2\rhalf\ is approximately proportional to $r_s$, being smaller by a factor of $\sim 2$, on average, but the difference decreases as the clusters are more massive. Therefore, according to the scenario discussed in \citet[][]{Chen+2022}, the radius 2\rhalf$\lesssim r_s$ could indicate the end of the BCG sphere of influence. By decomposing the stacked BCG+ICL light profile (after masking satellites) of $\sim 3000$ clusters at $0.2<z<0.3$ from the SDSS, these authors find that the inner region (BCG) is well fitted by a de Vaucouleours law, while the outer region is described by a projected NFW profile associated to the dark matter halo, which they infer from weak lensing constraints. The sum of both components evidences the need of a third intermediate component at the transitional scales between $\approx 70$ and 200 kpc for the stacked profile. Interestingly, this is of the order of the range of scales corresponding to 2\rhalf\ from our TNG300 cluster sample.

From the discussion and results presented above, it could be that 2\rhalf\ is an appropriate breakup radius to \textit{nominally} separate the ICL from the BCG. 
We have shown that by using this radius, the ICL fractions $\ficl$ and $\ficltot$ have no dependence or only a weak dependence on $\Mtwo$, respectively, and have the lowest dispersion among the different aperture-based criteria we examined (Fig.~\ref{fig:median_f_F_icl_vs_m200_diff_appertures}).
We have also seen that using the more theoretical radius $0.1\Rtwo$ to separate the ICL from the BCG implies ICL mass fractions ($\ficl$ and $\ficltot$) generally of the same order as those obtained with $r_{\rm ap}=$2\rhalf, but with much higher dispersion and slightly different trends with $\Mtwo$ (see Fig.~\ref{fig:median_f_F_icl_vs_m200_diff_appertures}).

Finally, it is interesting to note that \citet{proctor2024}, using a kinematic approach based on a Gaussian mixture model to identify the intra-halo light (IHL) in the EAGLE simulation, found that the mass associated with the IHL exhibits a 1:1 correlation with the mass enclosed within 2{\rhalf} for their most massive objects, which are comparable in scale to some of the clusters in our sample (see their Figure 6). Furthermore, the ICL fractions determined by \citet{proctor2024} for the simulations of EAGLE and Cluster-EAGLE, $\ficl\sim 0.3-0.4$, obtained for the clusters with $\Mtwo \geq 10^{13}$, are very similar to those we measure for TNG300 when using 2\rhalf\ to separate the ICL from the BCG, as will be shown below. These results provide a strong theoretical justification for opting 2\rhalf\ as a reliable criterion to (nominally) separate the ICL from the BCG.

\begin{figure}
  \centering
  \vbox{
    \includegraphics[width=\linewidth]{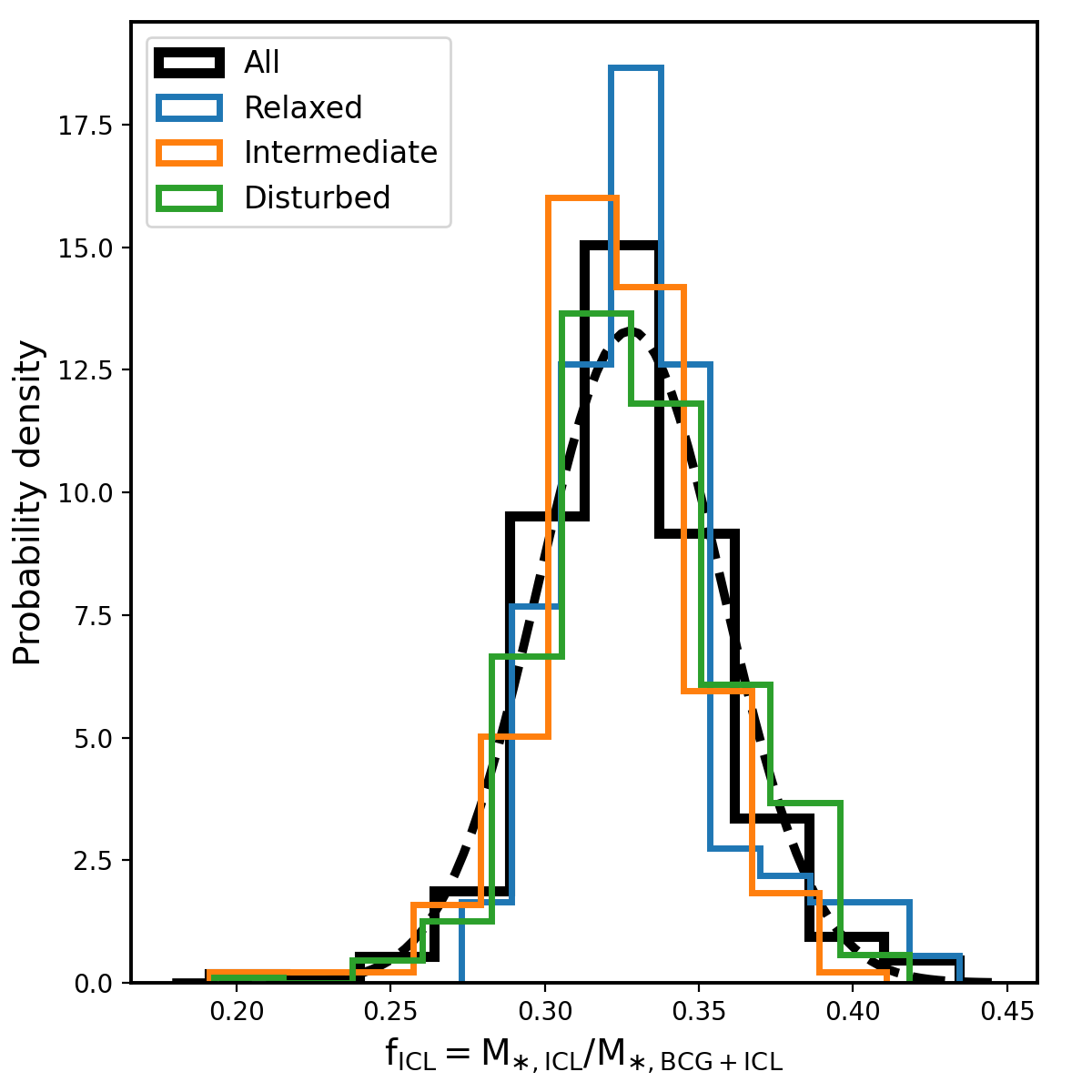}
    \\
    \includegraphics[width=\linewidth]{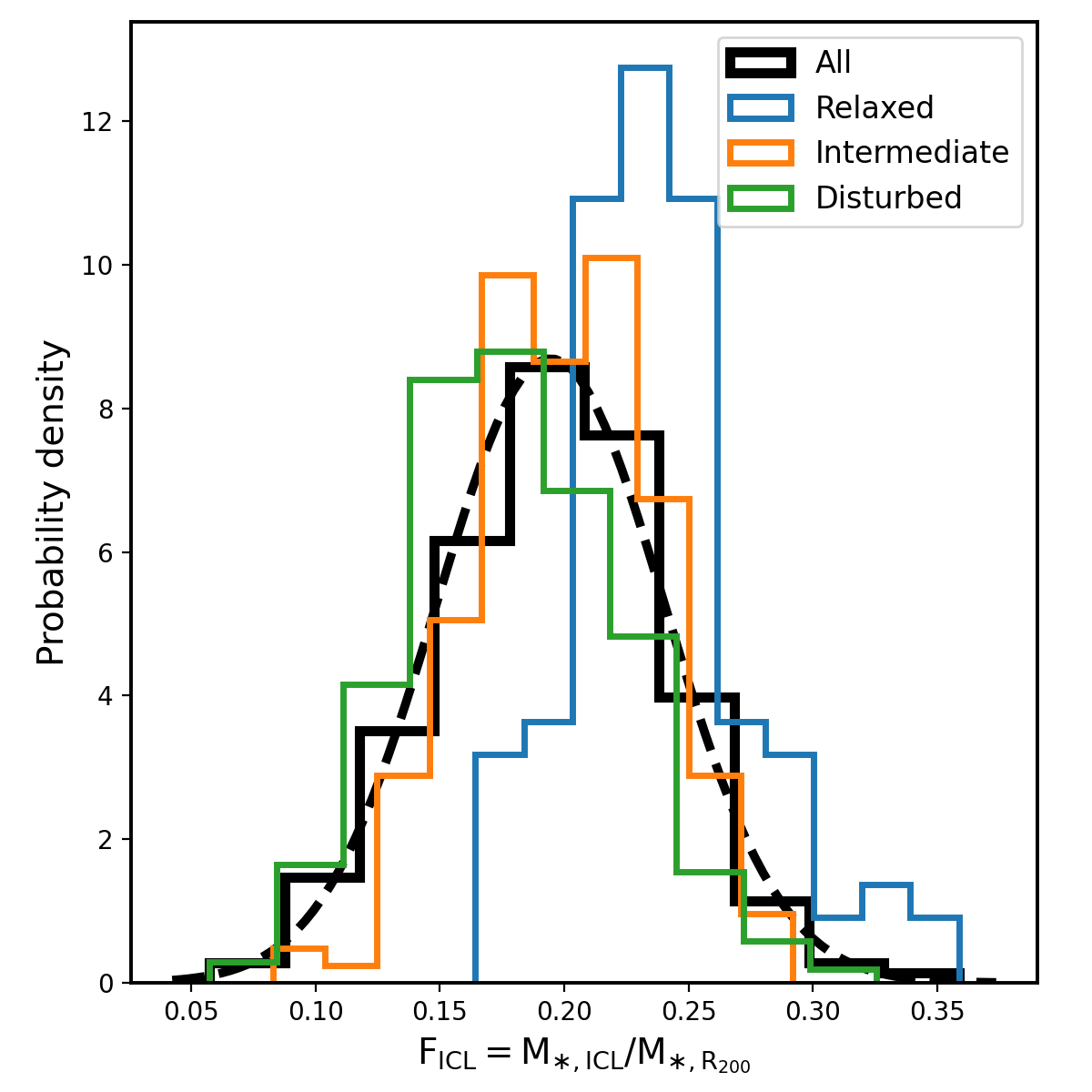}
    }
    \caption{Probability density of the ICL fraction using their two definitions: $\ficl$ (top panel) and $\ficltot$ (bottom panel). The solid black line shows the probability density for the whole sample and the dashed black line shows the fit to the distribution of a Gaussian function. We separate the probability distributions into three dynamic states: relaxed, intermediate and disturbed (solid blue, orange and green lines, respectively).}
	\label{fig:hist_f_F_ilc_dynamical_state}
\end{figure}

\begin{figure*}
  \centering
    \includegraphics[width=\linewidth]{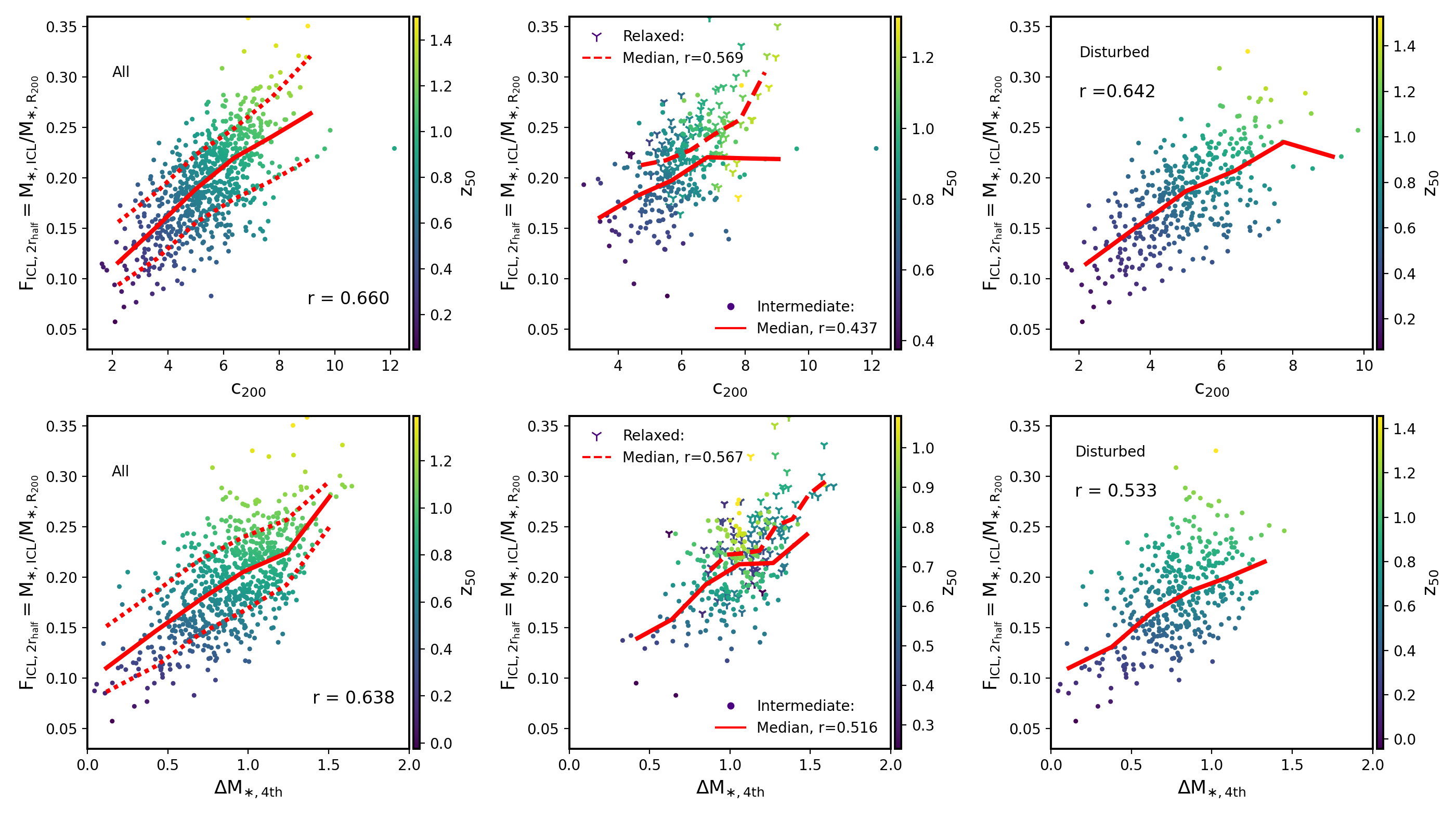}
	\caption{Total ICL mass fraction, $F_{\rm ICL}$ with respect to the total stellar mass (\mstartotal) as a function of concentration ($\ctwo$; top panels) and mass gap ($\deltam$; bottom panels). The colour scale in all panels represents the half-mass assembly redshift, $\zfive$. The colour of the data points has been smoothed using the LOESS algorithm. The left panels are for the full cluster sample, with the solid and dashed lines showing the medians and 16-84 percentiles, respectively. The middle panels are for the relaxed (skeletal triangles) and intermediate (dots) clusters only, and the right panels are for the disturbed clusters only.  The respective medians are indicated by the dashed and solid lines in the middle panels and by the solid lines in the right panels. The corresponding Pearson correlation coefficients, $r$, are shown within each panel.}
	\label{fig:scatter_F_icl_vs_c200-deltam4_cz50_dyn_state}
\end{figure*}

\subsection{What are the factors governing the ICL mass fractions?}
\label{sec:ICL-fraction-dependancies}

In Section~\ref{sec:BCG+ICL} we learned that the BCG+ICL (central object) stellar mass fraction, $F_{\rm BCG+ICL}=M_{\rm BCG+ICL}/\Mtwost$, is weakly anticorrelated with $\Mtwo$, but strongly correlated with $c_{200}$, $z_{\rm 50}$, and the gap $\deltam$. 
An interesting question is how the BCG and ICL components separately correlate with these properties. However, this depends on what criterion we use to separate the ICL from the BCG, as can be anticipated from  Fig.~\ref{fig:median_f_F_icl_vs_m200_diff_appertures}, at least in what regards the dependence on $\Mtwo$. Following, we will use the criterion $r_{\rm ap}=2$\rhalf\ (see previous subsection for some arguments in favor of this criterion) to explore how the ICL stellar mass fractions depend on these properties and on the cluster dynamical state. The results obtained will be  qualitatively similar to those using $r_{\rm ap}=0.1 \Rtwo$.

In the top and bottom panels of Fig.~\ref{fig:hist_f_F_ilc_dynamical_state} we present the distributions of $\ficl$ and $\ficltot$, respectively, for our TNG300 cluster sample (black solid histograms) using $r_{\rm ap}=2$\rhalf. The distribution (scatter) of $\ficl$ is very tight, with a mean and standard deviation of $\ficl=0.327\pm 0.029$. As for $\ficltot$, it presents a larger scatter, with $\ficltot=0.194\pm 0.045$.
In both cases, the distributions are well approximated by a Gaussian function (dashed lines). 

Using the criteria discussed in Sec.~\ref{sec:dynamical-stage} to classify the TNG300 clusters into three dynamical states, we obtain their respective distributions of  $\ficl$ and $\ficltot$ (colour-coded histograms). Within the small dispersion of $\ficl$ there is no significant segregation by dynamical state. The medians of $\ficl$ for the relaxed, intermediate and disturbed clusters are 0.33, 0.32, and 0.32, respectively. 
As for $\ficltot$, which takes into account the total stellar mass in the satellites, within its large scatter there is a clear segregation by dynamical state, with its value increasing the more relaxed the cluster is. The medians of $\ficltot$ for the relaxed, intermediate and disturbed clusters are 0.21, 0.16, and 0.13, respectively. Thus, what really seems to depend on the dynamical state is the fraction of total stellar mass in satellites,$F_{\rm \ast,sat}$ (or its complement, the mass fraction in the central object, $F_{\rm \ast,cen}$): the fraction of mass in the central object increases with the degree of relaxation because this also increases the time available for dynamical friction and tidal forces on the satellites -- but without a clear preference of transfer to the BCG or the ICL, since $\ficl$ does not depend on the dynamical state as shown in the top panel. 

We have also explored whether the ICL mass fractions correlate with $\ctwo$, $z_{\rm 50}$ or $\deltam$. For $\ficl$, we do not find any significant dependence. If anything, there is a very weak and noisy trend for $\ficl$ to decrease with $\deltam$. Instead, $\ficltot$ strongly correlates with these properties, as shown in the left panels of Fig.~\ref{fig:scatter_F_icl_vs_c200-deltam4_cz50_dyn_state}. The solid red lines indicate the median values, while the dotted lines represent the 16th and 84th percentiles. 
These results are similar to those shown for the BCG+ICL mass fraction $F_{\rm BCG+ICL}$ (see Figs. \ref{fig:scatter_f_bcg_icl_vs_m200-c200_cc200-cz50} and \ref{fig:scatter_f_bcg_icl_vs_z50-deltam4_cdeltam4-cz50}). The above implies that the BCG and ICL growth are affected approximately the same way by $\ctwo$ and depend similarly on $\deltam$. At fixed $z_{50}$, the more concentrated the cluster, the more efficient the loss of stellar particles from the satellites or their disruption in favour of both the BCG and ICL mass growth, and consequently the larger the mass gap. 

In the middle and right panels, we show the correlations with $\ctwo$ and $\deltam$ (coloured by $z_{\rm 50}$), separating the sample into relaxed (crosses)/intermediate (dots) and disturbed clusters, respectively. The solid lines indicate the medians of the intermediate and disturbed clusters, while the dashed lines indicate the medians of the relaxed clusters. The corresponding Pearson correlation coefficients are shown inside the panels. Relaxed clusters show higher $\ficltot$ fractions and a correlation of $\ficltot$ with $\ctwo$ stronger than intermediate and disturbed clusters. Regarding the correlation of $\ficltot$ with $\deltam$, the more relaxed the clusters, the stronger this correlation (and the higher the $\ficltot$ values). Relaxed clusters also tend to have larger mass gaps than intermediate and disturbed ones.

Despite that the scatter of $\ficltot$ is larger than that of $\ficl$, both show a normal distribution (see the Gaussian fits to the distributions in Fig.~\ref{fig:hist_f_F_ilc_dynamical_state}, shown as dashed lines). In the case of $\ficltot$, even for the relaxed, intermediate and disturbed cluster subgroups, although the distributions are shifted from each other, they also appear to follow distributions that are not very different from the normal. Using a semi-analytic prescription, \citet[][]{contini2023} also found that $\ficltot$ has a normal distribution. They then conclude that the processes that drive the formation (and evolution) of the ICL, such as stellar stripping, mergers and accretion of pre-processed diffuse stars, should have a stochastic nature. Recently, from the analysis of 11 clusters from the Horizon-AGN hydrodynamics simulation, \citet{brown2024} find that ICL assembly is stochastic because a significant source of ICL stellar particles is the tidal stripping of a small number of massive satellites (around $\sim 10^{11}$ \msun), which infrequently join the cluster. In the next Section, we present the stellar mass fractions in the ICL and BCG originating from different channels.

\begin{figure}
  \centering
  \vbox{
    \includegraphics[width=0.95\linewidth]{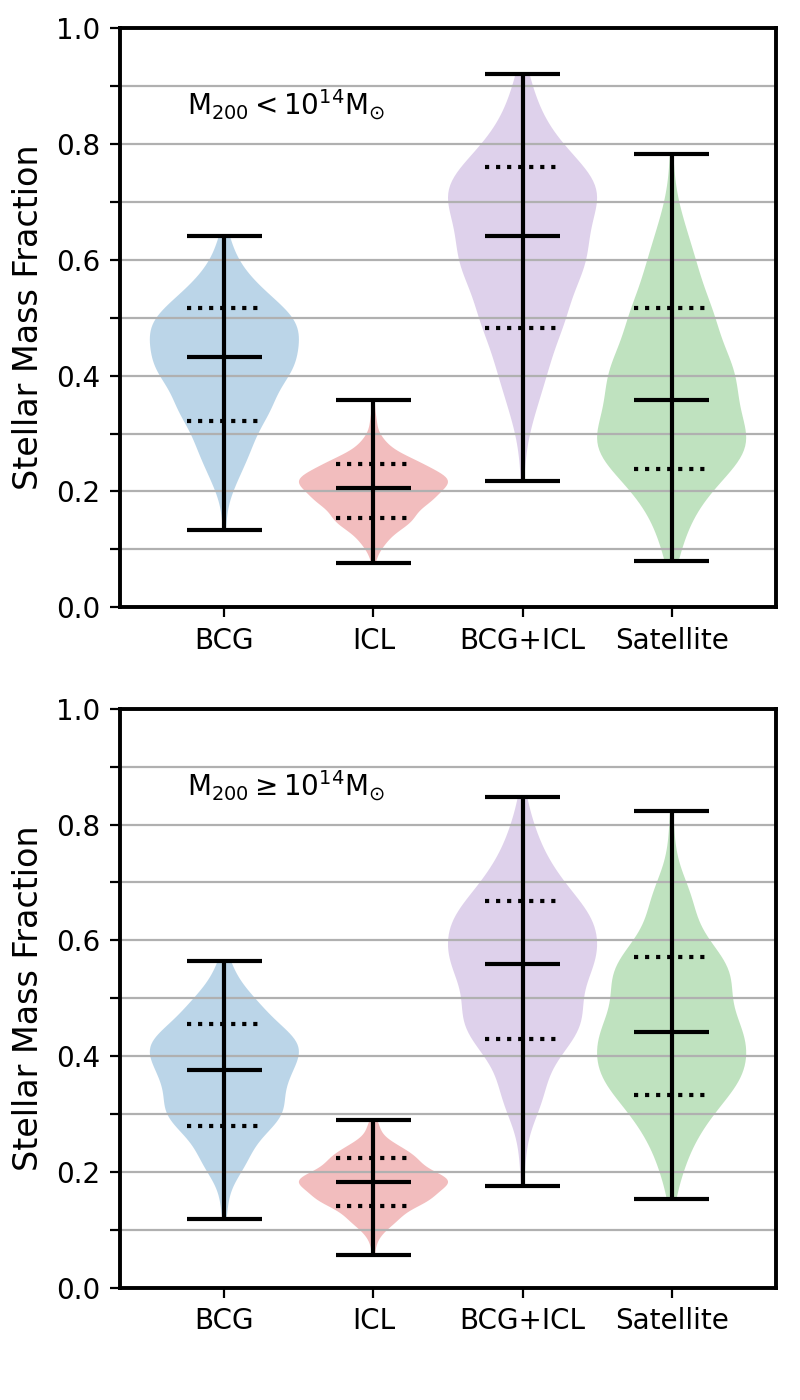}
    }
	\caption{Stellar mass fraction (with respect to $M_{\rm \ast,200}$) of the BCG, ICL, BCG+ICL, and Satellite galaxies for our sample of TNG300 clusters in the mass ranges $\Mtwo = 5 \times 10^{13} - 10^{14} \, \Msun$ (upper panel) and $\Mtwo > 10^{14} \, \Msun$ (bottom panel). The solid line represents the median, as well as the maximum and minimum values for each sample, while the dotted line marks the 16th and 84th percentiles. The BCG and ICL components are defined by the aperture radius 2\rhalf.}
    \label{fig:violin_plot}
\end{figure}

\section{Channels of BCG and ICL formation}

\label{sec:chanels}

Once we have introduced a way to \textit{nominally} separate the 3D central stellar object into BCG and ICL components using an aperture (transition) radius equal to 2\rhalf, we can measure the mass fractions at $z=0$ in each cluster stellar component and the fractions of the stellar particles in these components coming from different formation channels. 
The BCG stellar mass is the mass associated to the central object and contained within 2\rhalf, $M_{\rm \ast,BCG}=M_{\rm \ast,cen}(r < 2$\rhalf). The ICL stellar mass is the mass associated to the central object that lies beyond 2\rhalf\ and within $\Rtwo$, $M_{\rm \ast,ICL}=M_{\rm \ast,cen}$(2\rhalf$< r < \Rtwo$). 

In Fig.~\ref{fig:violin_plot}, we present different statistics, through violin plots, for the BCG, ICL, BCG+ICL, and Satellites total mass fractions, in two cluster mass ranges, $\Mtwo = 0.5$--$1 \times 10^{14} \, {\rm M}_{\odot}$ (top) and $\Mtwo \geq 10^{14} \, {\rm M}_{\odot}$ (bottom). The median central object mass fractions are $F_{\rm BCG+ICL}=0.64$ and $0.56$ for the low- and high-mass clusters, respectively, with wide probability density functions skewed towards lower values as shown with the purple shaded areas (see also the range of the 16--84th percentiles, dotted lines). The Satellites mass fractions are the complement of the BCG+ICL fractions (see also Figs. \ref{fig:scatter_f_bcg_icl_vs_m200-c200_cc200-cz50}-\ref{fig:scatter_f_bcg_icl_vs_z50-deltam4_cdeltam4-cz50}). The skewness in the $F_{\rm BCG+ICL}$ distribution is due to the BCG component, which has median values $F_{\rm BCG}=0.43$ and $0.37$ for the two mass ranges, and wide and skewed distributions. Regarding $\ficltot$, its distribution is close to a normal one, as discussed above, more for the high-mass clusters. The median values of $\ficltot$ are 0.21 and 0.18 for the low and high mass ranges, respectively.

\begin{figure}
  \centering
    \includegraphics[width=\linewidth]{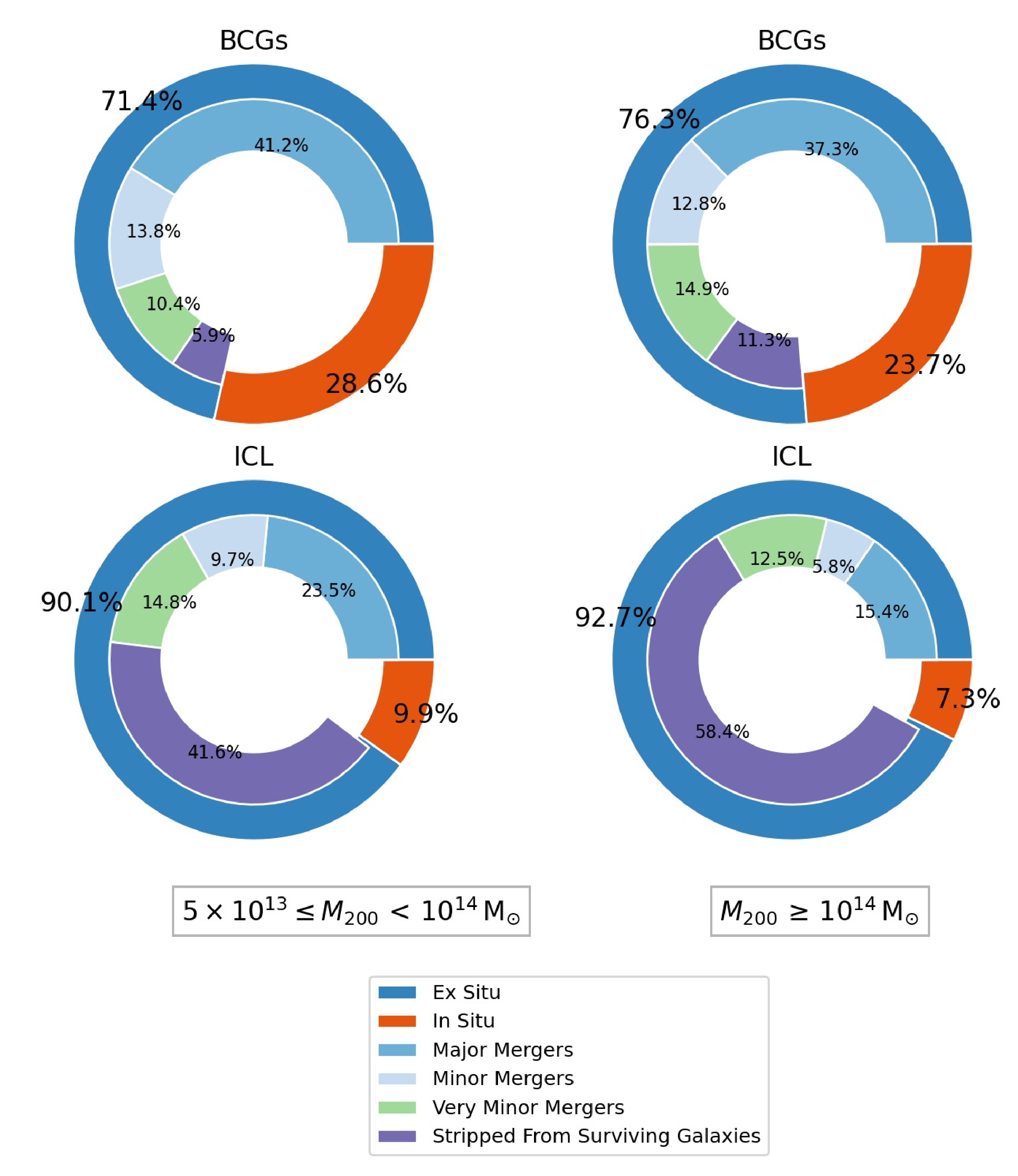}
	\caption{Means of the stellar mass fractions originating in different channels for the BCG (top) and ICL (bottom) components of our sample of TNG300 clusters in the mass ranges $\Mtwo = 5 \times 10^{13} - 10^{14} \, \Msun$ (left panels) and $\Mtwo = 10^{14-15.2} \, \Msun$ (right panels). In each pie chart, the outer layer shows how the stellar mass is split into \textit{in situ} (orange) and \textit{ex situ} (blue) components, while the inner layer separates the \textit{ex situ} stellar mass into the following components: stars accreted from (i) major mergers, (ii) minor mergers, (iii) very minor mergers, and (iv) stripped from surviving galaxies, as indicated in the legend. Note that the fractions in the second layer are normalised with respect to the \textit{total} stellar mass (that is, including both the \textit{in situ} and \textit{ex situ} components).
 }

	\label{fig:pie_charts_2samples_halo_mass}
\end{figure}

The two general channels of BCG and ICL stellar mass growth are by  \textit{in situ} star formation and \textit{ex situ} accretion of stellar particles.\footnote{A stellar particle is classified as \textit{in situ} if it formed within the galaxy or central object that is currently located along the main branch of its evolutionary history, and it is classified as \textit{ex situ} if it originated in a different galaxy and was subsequently acquired by the current galaxy or central object through mergers (disruption) or tidal stripping.} 
In Fig.~\ref{fig:pie_charts_2samples_halo_mass} we present the mean mass contributions from both channels for the BCG and ICL components, first layer of the pie charts. The second layer shows the mean ex-situ contributions to these components: major, minor and very minor mergers, and tidally stripped stars from surviving galaxies (satellites). 
A similar analysis was presented in \citetalias{montenegro2023} for the same sample of TNG300 clusters, but in that paper a fixed breakup radius of 30 kpc was used to separate the BCG from the ICL within the central object.

In order to follow the main formation channels for the ex-situ (accreted) component of the BCG+ICL (central object), we make use of the stellar assembly catalog described in \cite{Rodriguez-Gomez2016}, which provides us with information on the origin of each stellar particle within the simulation. The formation channels for stellar particles associated with accretion processes (ex-situ channel) presented in that work are labeled as ``mergers'' or ``stripping''; the mergers label represents stellar particles whose progenitor has already been fully accreted (disrupted) in a merger event with the central object, as \textsc{Subfind} defines it, while the stripping label refers to stellar particles whose progenitor currently survives, but transferred a fraction of its particles to the central object\footnote{Note that galaxies and groups have stellar haloes, whose stellar particles should be the first to be stripped by tidal forces. The contribution to the BCG+ICL from these halo stripped stellar particles is associated with the channel known in the literature as ``preprocessing''.}. The stellar mass ratios used to define major, minor, and very minor mergers are $\mu \geq 0.25$, $0.1\leq \mu < 0.25$, and $\mu<0.1$, respectively. The stellar mass ratio is defined as $\mu = M_{\ast, 2} / M_{\ast, 1}$, where $M_{\ast, 1}$ and $M_{\ast, 2}$ are the stellar masses of the primary and secondary progenitors measured at the moment when the secondary reached its maximum stellar mass \citep{Rodriguez-Gomez2015}. In the context of the current paper, the primary mass is that of the central object, $M_{\rm \ast,cen}$, and after the merger, all the secondary stellar particles belong to the primary, i.e., the secondary galaxy disrupts.

As seen in the outer layers of Fig.~\ref{fig:pie_charts_2samples_halo_mass}, the stellar particles in the BCG and ICL components were predominantly accreted, i.e., they are ex situ, with a slight difference in the fractions according to the cluster mass range. However, the ex-situ channel is much more relevant for the ICL than for the BCG.  For the low-mass sample, the mean ex-situ stellar mass fraction and its standard deviation are $71.4\pm12.8$ per cent in the BCG, while in the ICL are $90.1\pm3.8$ per cent (the complements of these fractions correspond to the in-situ stellar particles). For the high-mass sample, these mean ex-situ stellar mass fractions are only slightly larger, $76.3\pm9.7$ per cent for the BCG, while $92.7\pm3.4$ per cent for the ICL. Therefore, these fractions are weakly dependent on cluster mass. Note that for the ICL component, the dispersion of the ex-situ mass fraction is small, so that the reported mean fractions correspond roughly to the fractions measured individually for each cluster. Regarding the BCG component, the dispersion is large and skewed toward lower values. This points to a larger variation for BCGs in the contribution of the ex-situ channels than for the ICL. Indeed, as we discuss below, different mergers, especially major ones, dominate the ex-situ BCG formation channel, but the occurrence of major mergers is stochastic and infrequent.

\subsection{Channels of the ex-situ component}

Analyzing the formation channels of the ex-situ component (second layer of the pie charts), we find that in the low-mass range, mergers (major, minor, and very minor) strongly dominate, on average, over the stellar particles stripped from surviving galaxies in the BCGs (91.7 per cent vs. 8.3 per cent of the ex-situ stellar particles, respectively, major mergers being the dominant channel, albeit with a very wide distribution). 
At the same time, for the ICL, the channel of stripped stellar particles dominates over the others, with 46.2 per cent of the ex-situ stellar particles, on average, followed by major mergers with 26.1 per cent, very minor mergers with 16.4 per cent, and minor mergers with 10.8 per cent. 
Therefore, for both the BCG and the ICL in the low-mass clusters, all mergers contribute more to the ex-situ stellar particles than those particles stripped from surviving galaxies, being they much more relevant (in particular major mergers) for the BCG than for the ICL. 
In the high-mass range, the contribution to ex-situ particles in the BCG is, on average, 14.8 per cent from stripping of surviving satellites and 85.2 per cent from mergers, which is similar to the low-mass subsample. This changes dramatically, however, when we analyze the ICL, where the ex-situ mass is dominated by stripped stars from surviving galaxies at 63 per cent on average. 

\begin{figure}
  \centering
 \vbox{
    \includegraphics[width=\linewidth]{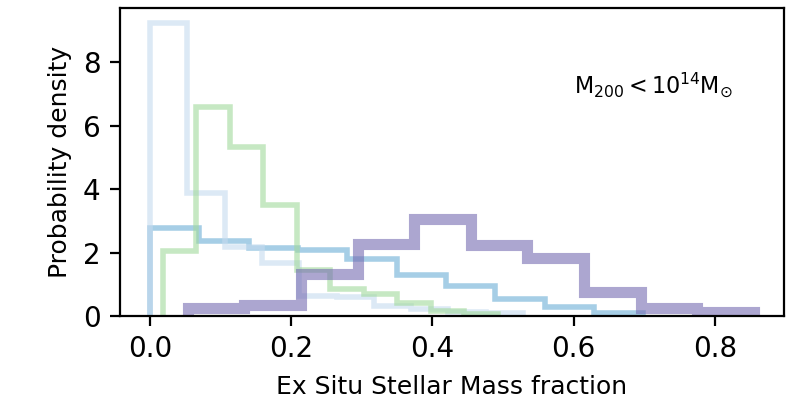}
    \\
    \includegraphics[width=\linewidth]{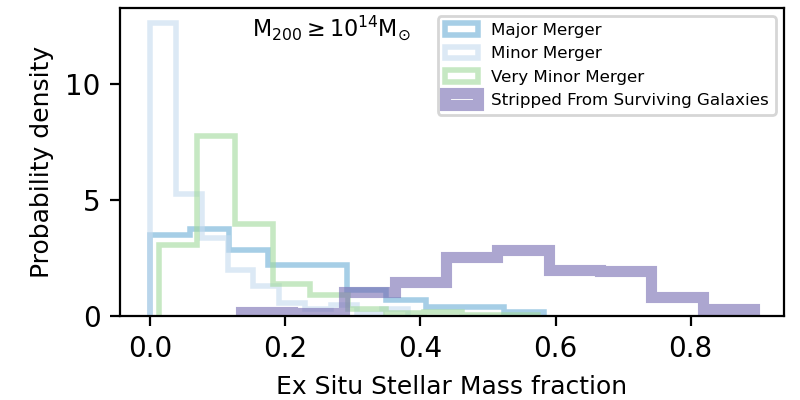}
    }
	\caption{Probability density of the ICL ex-situ stellar mass fraction for all TNG300 clusters within two different masses ranges: $\Mtwo = 5 \times 10^{13} - 10^{14} \, \Msun$ (upper panel) and $\Mtwo \geq 10^{14} \, \Msun$ (bottom panel). The ex-situ stellar mass is further categorized into different accretion channels: major mergers, minor mergers, very minor mergers, and stars stripped from surviving galaxies, as indicated in the figure labels.}
	\label{fig:Histogram_channeles_formation}
\end{figure}

In Fig.~\ref{fig:Histogram_channeles_formation}, we show the distributions of the ICL mass fractions contributed by the different ex-situ channels in both mass ranges.  The stellar stripping channel, which is the dominant, presents a bell-shaped distribution, very wide, more in the high-mass clusters, where it can be fitted by a normal distribution. The second channel contributing to the ex-situ stellar particles of the ICL, major mergers, has a positively skewed distribution, with its peak at $\lesssim 5$ per cent, i.e., the frequency of major mergers contribution to the ICL ex-situ component decreases as larger is this contribution. On the other hand, very minor mergers (i.e. the disruption of low-mass galaxies), have a more constant contribution to the ICL around 10--15 per cent.

In summary, we see that the main difference in the channels of ex-situ stellar mass growth between the BCG and ICL components as defined here is in the fractional contribution of stellar particles from tidal stripping of surviving galaxies, and that from mergers. Both channels contribute similarly to the ICL, if anything, the former channel dominates in more massive clusters, while the latter channel strongly dominates for the BCG, with major mergers being the main source of their ex-situ growth. This is consistent with the theoretical expectation that the satellite galaxies lose substantial orbital angular momentum due to dynamical friction, which drives them to the cluster's centre, where they may undergo mergers with the central galaxy (BCG) and increase the BCG's mass -- specially via major mergers, which are infrequent and stochastic. 

By comparing the results of this study with those of \citetalias{montenegro2023}, 
where an aperture radius of 30 kpc was used to distinguish between the BCG ($r < 30$ kpc) and the ICL ($30$ kpc $< r < \rm{ R_{500}}$), 
we observe no significant overall differences in the mean in-situ and ex-situ mass fractions of the BCG and ICL components for the two mass subsamples. For example, for the ICL component, the  mean ex-situ mass fraction reported here and in \citetalias{montenegro2023} are 0.901 vs. 0.871 (0.927 vs. 0.886) in the low-mass (high-mass) regimes, respectively.
Since 2\rhalf\ is larger than 30 kpc in almost all clusters, slightly larger ex-situ mass fractions in the ICL are captured here because the ex-situ component dominates more at larger clustercentric radii.
Moreover, the ex-situ component distribution also shifts, particularly within the ICL, transitioning from being predominantly influenced by stellar particles accreted through mergers to being primarily shaped by tidal stripping. 

It is interesting to note that there is a non-negligible amount of in-situ stellar particles in the ICL (9.9 per cent and 7.3 per cent for low- and high-mass clusters, respectively, see Fig.\ref{fig:pie_charts_2samples_halo_mass}). Although this is a small amount, it is challenging to explain how star formation could have occurred directly in the ICL (see for a discussion, \citetalias{montenegro2023}). 
\cite{ahvazi2024_insitu_ICL} have found also a non-negligibly contribution of in-situ stars to the ICL; they estimated from 8 to 28 per cent for the three most massive clusters in the TNG50 simulation. Their study shows that the majority of these in-situ stellar particles were formed within the ICL itself, in cold clouds that condense along large filaments (not related to gas lost by galaxies), hundreds of kiloparsecs from the central galaxy. Contrary to our results and those of \cite{ahvazi2024_insitu_ICL}, \citet{brown2024} find that in-situ star formation in the 11 Horizon-AGN clusters they analyzed is negligible. Whether or not star formation can occur in situ in the intracluster medium is an open question.

\subsection{The fractions for different dynamical states}
\label{sec:dynamical-states}
Do the different channels of BCG and ICL growth depend on the dynamical state of the cluster? To answer this question, we have calculated the different fractions presented in Fig.~\ref{fig:pie_charts_2samples_halo_mass}, but now for the disturbed, intermediate and relaxed clusters, also separating them into low- and high-mass groups. The resulting pie charts with the mean values of the fractions are presented in the Appendix, Fig.~\ref{fig:pie_charts_dyn_state_samples}.  According to this figure, the fractions of in-situ/ex-situ stellar particles in the BCG and ICL components are roughly the same for the clusters independent of their dynamical state. If anything, the ex-situ (in-situ) mass fraction is slightly lower (higher) in the disturbed clusters than in the relaxed ones for both the BCG and ICL components. As for the ex-situ component in the ICL, the contribution of tidally stripped stellar particles from surviving galaxies is slightly larger in the disturbed clusters than in the relaxed ones.

\section{Discussion}
\label{sec:discussion}
\subsection{Implications of our results}
For the TNG300 clusters analyzed here, we have found that the fraction of stellar particles in the central object (BCG+ICL) relative to all stellar particles within $\Rtwo$, $F_{\rm BCG+ICL}$ (see Table \ref{tab: Definition components}), increases strongly with the cluster NFW concentration (or the half-mass assembly redshift, $z_{50}$; see Section~\ref{sec:BCG+ICL}). Since tidal forces are stronger in more concentrated clusters, more efficient tidal stripping and disruption (mergers) of satellite galaxies is expected in these clusters, with the stellar particles ending up in the BCG+ICL component. 
This is indeed suggested by the top panel of Fig.~\ref{fig:scatter_f_bcg_icl_vs_m200-c200_cc200-cz50}, where the $F_{\rm BCG+ICL}$ increases with $\ctwo$ for a fixed $\Mtwo$. Furthermore, if the cluster was assembled earlier, there is more time for this channel of ex-situ growth to act. We have also shown that the ICL grows more through the ex-situ channel than the BCG (Section~\ref{sec:chanels} and Fig.~\ref{fig:pie_charts_2samples_halo_mass}), in such a way that $\ficltot=M_{\rm \ast,ICL}/M_{\rm \ast,200}$ also strongly correlates with $\ctwo$ (and $z_{50}$), a result also reported and discussed in \citet{contini2023}.  Moreover, $\ficltot$ tends to be higher in more relaxed (evolved) clusters. A similar result was reported in \citet{contreras2024} for massive clusters from ``The Three Hundred'' hydrodynamics simulations. These authors also found a strong correlation of $\ficltot$ with $z_{50}$ (they used $r_{\rm ap}=50$ kpc to separate the ICL from the BCG).

For the criterion used here to separate the ICL from the BCG, an aperture radius $r_{\rm ap}=2$\rhalf, the obtained ICL mass fractions, $\ficl\approx 0.32$, have a narrow and close to normal distribution and do not depend on any cluster property. This is because the variations in the radial mass distribution of the BCG+ICL as a function of \rhalf\ are small and stochastic. 
If $\ficl$ is nearly constant but $\ficltot$ has a broader distribution with a strong dependence on $\ctwo$, $z_{50}$, and the dynamical state, then what depends on these properties is actually the fraction of mass in satellites, $F_{\rm \ast,sat}$ (or its complement $F_{\rm \ast,cen}$), given that $\ficltot=\ficl\times F_{\rm \ast,cen}= \ficl\times (1-F_{\rm \ast,sat})$. 
As mentioned above, the total stellar mass in the satellites seems to decrease due to tidal stripping and their disruption (more as more concentrated and relaxed is the cluster) in favour of the central object, but without a clear preference for the transfer of stellar particles to the BCG or the ICL, a process that seems to be more stochastic.  What changes significantly between the BCG and the ICL is the contribution of the different ex-situ channels: in the BCGs, major mergers dominate, with the smallest contribution from tidal stripping, whereas in the ICL, tidal stripping of surviving galaxies is the main channel \citep[see also][]{cooper2015,Contini2018,Contini2021,Chun+2023}.

As in many previous observational works \citep[e.g.,][and more references therein]{gonzalez2005,Seigar2007,donzelli2011, Kravtsov2018,zhang2019,Kluge+2020,montes2022Nature} and theoretical studies \citep[e.g.,][]{cui2014,cooper2015,Pillepich2018a,Contini2021,Brough+2024}, our analysis shows the difficulty of defining a transitional radius between the BCG and the ICL due to the smooth stellar mass (or light) distribution between the two components.  
For reasons discussed in Sect. \ref{sec:optimal-radius} and from comparisons of our ICL fractions with those from hydrodynamical simulations, where the stars bound to galaxies and those remaining in the ICL have been determined using dynamical approaches (see subsect. \ref{sec:ICL-comparisons}), 2 effective radii of the BCG+ICL component may be a good approximate radius to separate the ICL from the BCG. 
From the observational point of view, it is straightforward to determine this radius once the satellite galaxies and other structures are masked or subtracted from the cluster image. In this case, the analysis is performed for projected quantities and for light rather than stellar mass. In Section~\ref{sec:projected} we show that our results are very similar if the analysis is performed for projected distributions. The change from mass to light is expected to introduce minor corrections in the BCG and ICL fractions. 

\subsection{Proxies of $\ctwo$ and $\Mtwo$ for galaxy clusters}\label{proxy}
\begin{figure}
  \centering
    \includegraphics[width=\linewidth]{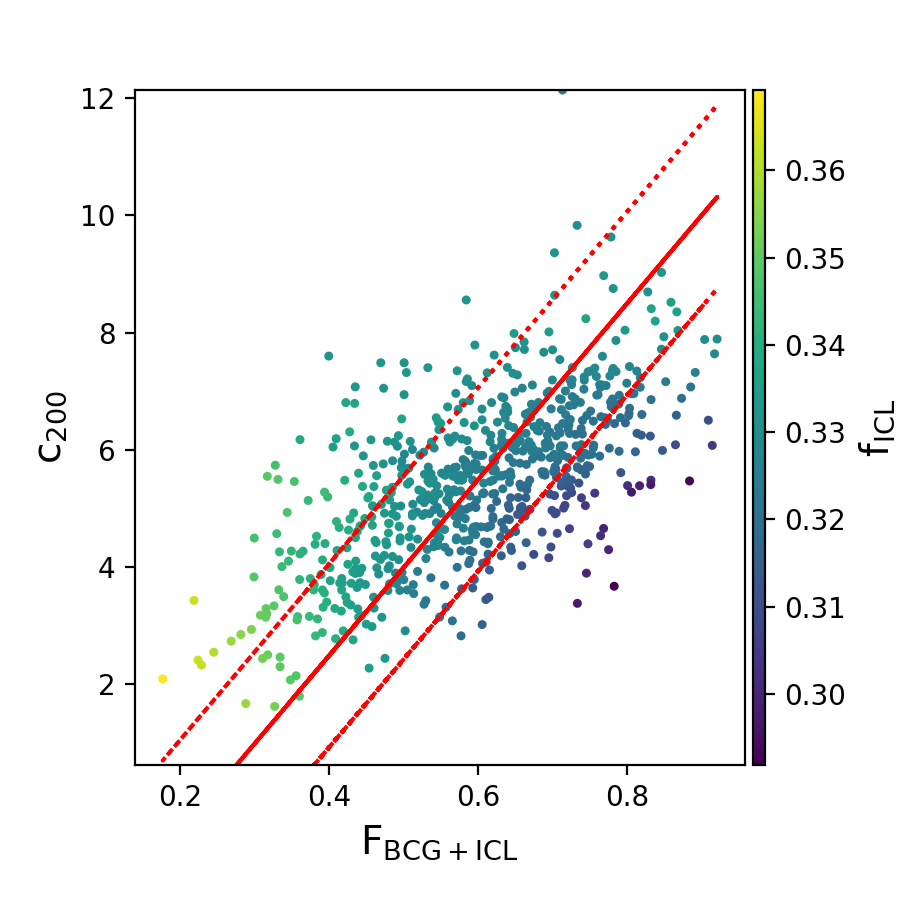}
	\caption{The $\ctwo$--$F_{\rm BCG+ICL}$ scatter plot along with the maximum-likelihood linear fit 
 [equation~(\ref{c200predict}); solid red line]. The dashed lines represent the $1$-$\sigma$ scatter about the relation. The colour scale represents the $\ficl$ fraction, and the colour of the data points has been slightly smoothed using the LOESS algorithm. This relationship [or the one given by equation~(\ref{c200predictwcor})] can be applied to observed clusters to estimate the halo concentration.}
	\label{fig:c200_vs_F_bcgpicl}
\end{figure}

Correlations between stellar and halo characteristics, if strong enough, can be leveraged to provide
predictions for the latter with a reasonable degree of accuracy. We have found that the $F_{\rm ICL}$ and $F_{\rm BCG+ICL}$ mass fractions correlate with $\ctwo$. In the opposite direction, by examining the correlations of $\ctwo$ with various stellar properties, we have found that the strongest trends are precisely with $F_{\rm ICL}$ and $F_{\rm BCG+ICL}$, and at similar strengths. However, $F_{\rm BCG+ICL}$ is relatively easier to quantify from observations (actually, its light and projected version) because it avoids the complications associated with separating ICL from combined emission. This makes $F_{\rm BCG+ICL}$ a preferable predictor of $\ctwo$ out of the two. Furthermore, we find that the Pearson and Spearman rank coefficients for the $c_{200}$--$F_{\rm BCG+ICL}$ relation are nearly the same, thereby implying that it can be well-approximated by a linear relation.
We obtain the linear fit by assuming the uncertainties in $c_{200}$ and $F_{\rm BCG+ICL}$ as zero, and maximising the log-likelihood proposed by \citet{Robotham2015} [equation~(13) in their paper]:
\begin{equation}\label{likelihood}
\ln{\mathcal{L}} = 0.5\sum_{i=1}^{N}\left[\ln{\frac{a^2+1}{\sigma^2}-\left(\frac{aF_{{\rm BCG+ICL},i}+b-c_{200,i}}{\sigma}\right)^2} \right],    
\end{equation}
where $N$ is the sample size, $a$ is the slope, $b$ is the intercept, and $\sigma$ is the
(global) vertical scatter about the line. The best-fit is shown in Fig.~\ref{fig:c200_vs_F_bcgpicl} and given by
\begin{equation}\label{c200predict}
    \ctwo = 15.022 F_{\rm BCG+ICL} - 3.514\,,
\end{equation}
with $\sigma=1.565$. The posterior distributions for $a$, $b$, and $\sigma$ are provided
in Appendix~\ref{posterior}.

The accuracy of the fit can be improved by adding the contribution of parameters that correlate with the vertical offset about the relation. Examining these correlations, we find that $F_{\rm BCG+ICL}$ has the strongest trend with the offset. The reason for this is the asymmetric distribution of points around the fit at the lowest and highest values of $F_{\rm BCG+ICL}$, resulting in a negative correlation between the offset and $F_{\rm BCG+ICL}$. This implies that if a stellar parameter is well correlated with $F_{\rm BCG+ICL}$, its trend with the offset may reflect the offset--$F_{\rm BCG+ICL}$ trend rather than the variation of the parameter with $c_{200}$ at a fixed $F_{\rm BCG+ICL}$. Hence, we carry out partial Spearman rank tests to extract the trends between the parameters and the offset after accounting for their covariances with $F_{\rm BCG+ICL}$. The results reveal that $\deltam$, \rhalf$/\Rtwo$, $\ficltot$, $\ficl$ are correlated with the offset at $>3\sigma$ level and correlation coefficients $\gtrsim 0.15$, in decreasing order of strengths.

We keep the slope and intercept fixed to those in equation~(\ref{c200predict}), add each of the above five parameters through a multiplicative factor, and determine the factor's value as the one corresponding to the least-squares. This is repeated for all the combinations of these parameters, and for each case, we use the Levene test \citep{Brown1974} to assess the significance of the changes induced in the variance of the offsets, and the Shapiro-Wilk test \citep{Shapiro1965} to assess normality. The idea is that these corrections should result in offsets with lower scatter and devoid of any global bias -- that is, the $\sigma$ indeed represents symmetric errors. Based on this, we propose a combination of $\deltam$ and $\ficl$ to correct for the scatter. The relation after incorporating the parameters is,
\begin{equation}\label{c200predictwcor}
        \ctwo = 15.022 F_{\rm BCG+ICL} - 3.514 - 2.466\deltam + 6.730\ficl,
\end{equation}
and carries the $\ctwo$ uncertainty of $1.380$, which is an improvement over the
default relation by $\approx 12$ per cent.

\begin{figure}
  \centering
    \includegraphics[width=\linewidth]{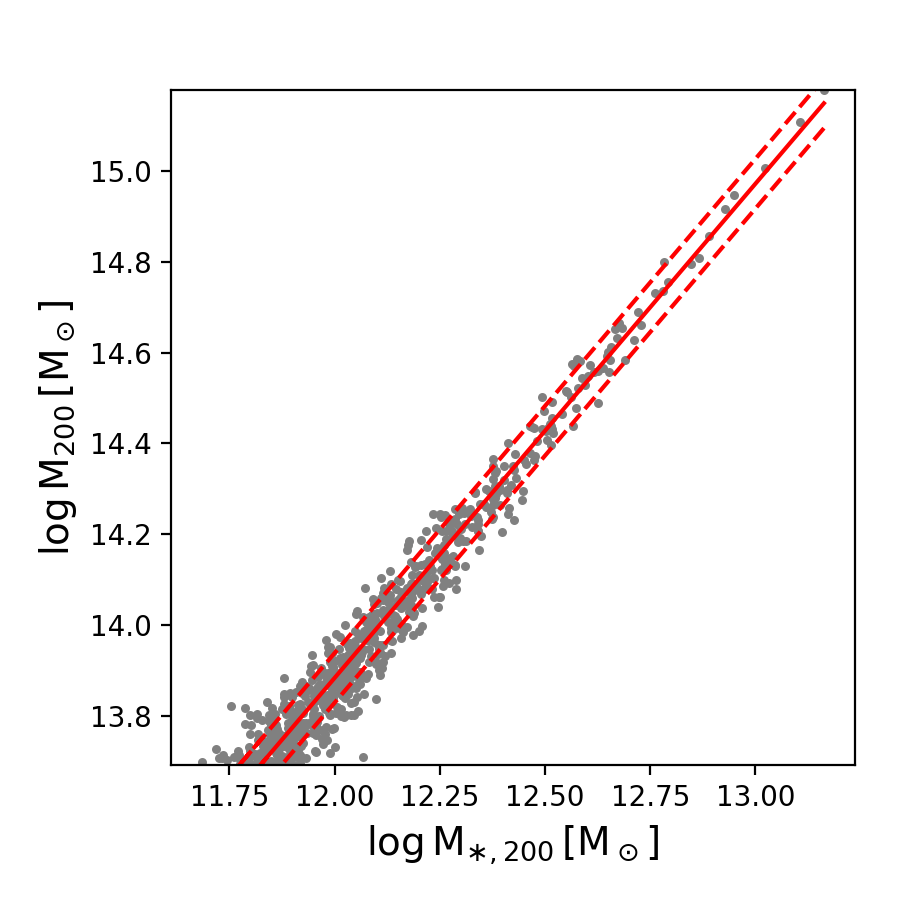}
	\caption{Cluster mass ($\Mtwo$) plotted against the total stellar mass ($\Mtwost$) along with the maximum-likelihood linear fit and the $1$-$\sigma$ uncertainty [equation~(\ref{m200predict})].}
	\label{fig:m200_vs_mstar}
\end{figure}

In the same spirit as for $\ctwo$, we try to identify a proxy for the cluster mass $\Mtwo$ and find the strongest correlation with $\Mtwost$ (Pearson coefficient of 0.99). 
The maximum-likelihood linear fit (shown in Fig.~\ref{fig:m200_vs_mstar}) describing the 
$\Mtwo$--$\Mtwost$ relation is given by
\begin{equation}\label{m200predict}
    \log\Mtwo = 1.089 \log \Mtwost + 0.815\,,
\end{equation}
with an uncertainty of $0.055$~dex. We explored various parameters to correct for the scatter in the relation using a similar approach to $\ctwo$, and found that additional parameters have a negligible impact on its predictability. We only analyze the quantities related to the aperture radius 2\rhalf, but we have seen that similar qualitative results are obtained for other apertures.

\subsection{How much do ICL mass fractions change from intrinsic to projected distributions?}
\label{sec:projected}

As stated in Sec. \ref{sec: Intro}, our goal in this paper was to find \textit{intrinsic} BCG+ICL and ICL mass fractions and their dependences on cluster properties, i.e. directly from the spherical-averaged (3D) stellar mass distributions. As a first step to explore how our results might connect to observations, we now explore how much these results change when we move to projected (2D) mass distributions. We only analyze the quantities related to the aperture radius 2\rhalf, but we have seen that similar qualitative results are obtained for other apertures. 

As expected 2\rhalfpro\ is always smaller than 2\rhalf. The difference is of $26\%$ on average, with a 1-$\sigma$ scatter of $\pm 8\%$. We do not observe any trend with $\Mtwo$ when comparing the difference between the projected and intrinsic (3D) radii, even when analyzing the different dynamical states. 

\begin{figure}
  \centering
  \vbox{
    \includegraphics[width=1.05\linewidth]{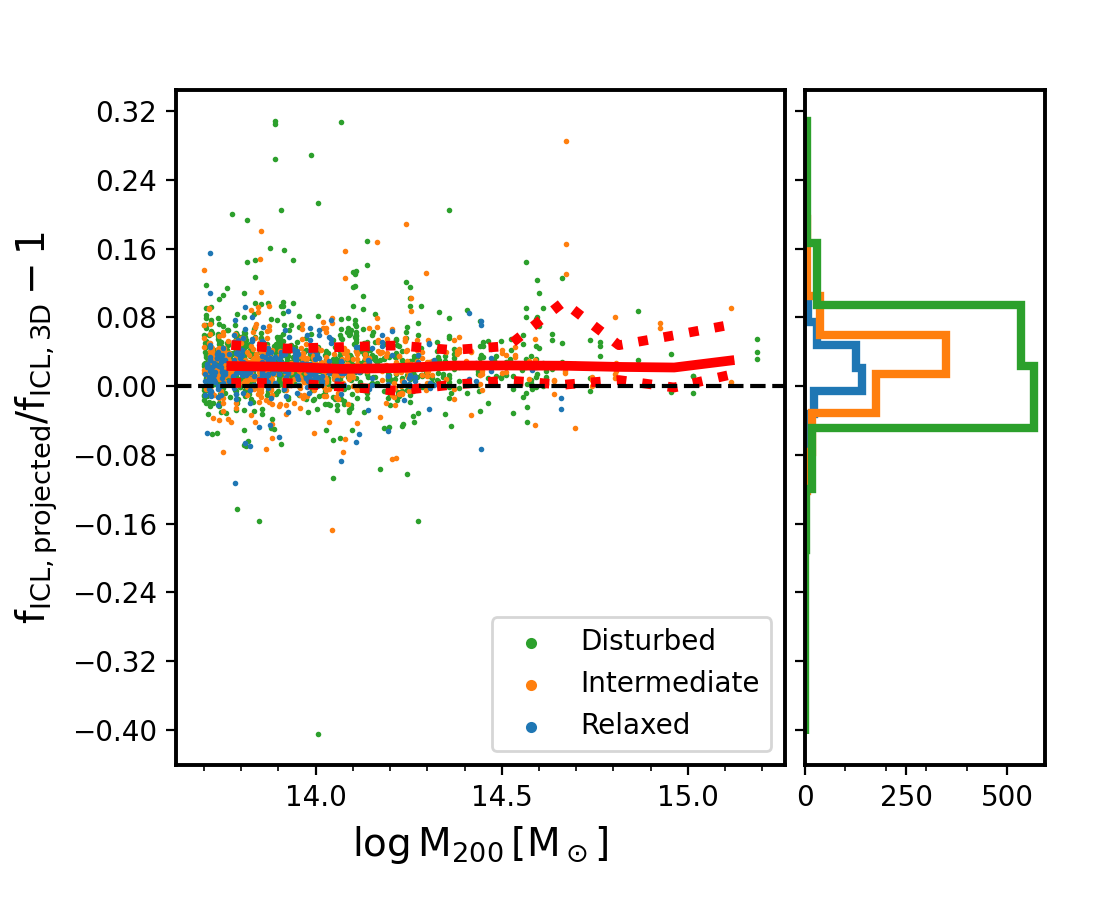}
	\\
    \includegraphics[width=1.05\linewidth]{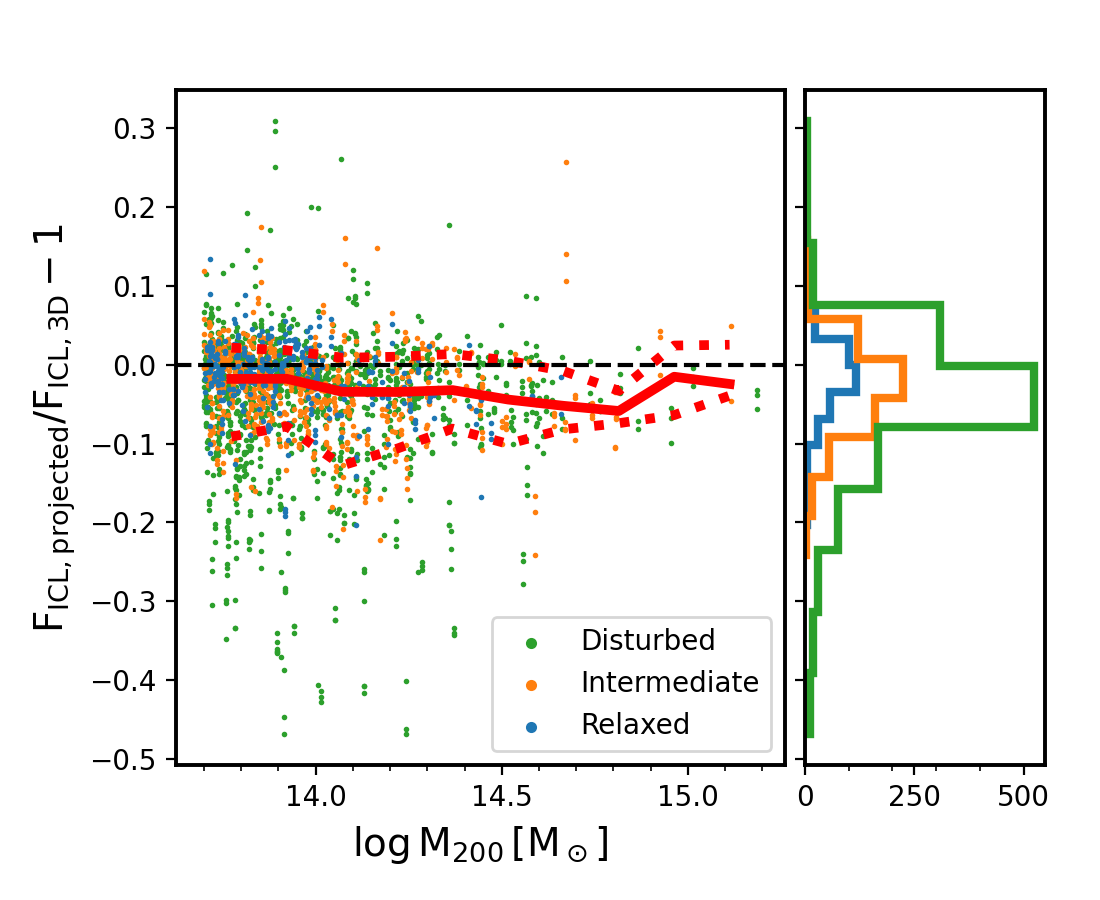}
    }
    \caption{Fractional ratios of the ICL mass fraction between projected distribution (2D) and intrinsic distribution (3D) are presented as a function of $\Mtwo$. We have projected each cluster in three orthogonal directions. The upper panel shows the fractional ratio for $\ficlpro/\ficl$, while the bottom panel displays $\ficltotpro/\ficltot$. For the entire sample, the solid red line represents the median values, while the dotted red lines indicate the 16th and 84th percentiles. The data points are color-coded based on the dynamical state of the clusters: disturbed (green), intermediate (orange), and relaxed (blue). In the right panel, the corresponding histogram of fractional ratios is plotted for each dynamical state.}
	\label{fig:median_f_F_icl_2D_3D-1_vs_m200_dyn_state}
\end{figure}

In Fig.~\ref{fig:median_f_F_icl_2D_3D-1_vs_m200_dyn_state} we show the fractional ratios of the ICL mass fractions $\ficl$ and $F_{\rm ICL}$ from projected distributions to intrinsic (3D) ones, respectively $\ficlpro/\ficl$ and $\ficltotpro/\ficltot$, as a function of $\Mtwo$.  To increase the statistics, we have projected our cluster sample in three orthogonal directions,  X--Y, X--Z, and Y--Z, using the natural coordinate system of the simulation. In Fig.~\ref{fig:median_f_F_icl_2D_3D-1_vs_m200_dyn_state}, the data is colour-plotted by the dynamical state of the cluster, and the solid and dotted lines show the median 16-84th percentiles, respectively. The differences between the projected and intrinsic ICL fractions are small for both $\ficl$ and $\ficltot$. For $\ficl$ ($\ficltot$), the median fractions obtained from projected distributions are $\sim 2\%$ (-3\%) greater and smaller respectively than the intrinsic ones, with a 1-$\sigma$ scatter of $\sim$ $\pm 4\%$ ($\pm 8\%$). While for $\ficltot$ there is a weak tendency for the difference to decrease as the clusters become more massive, for $\ficl$, this tendency is not observed. On the other hand, we do not observe that the dynamical state influences the differences. If anything, for $\ficltot$, the largest differences found in our analysis correspond to disturbed clusters. 

In summary, the main results presented in Sections \ref{sec:results} and \ref{sec:chanels} remain roughly the same when analyzing projected distributions, as is done for observations. However, the observational analysis is based on light distributions, i.e. photometric images. Because clusters generally exhibit negative M/L gradients, it is possible that the ICL fractions are slightly larger in light than in mass. Furthermore, observational effects on the images such as sky background and PSF have to be taken into account. In a companion paper (Montengro-Taborda et al., in prep) we generate mock images from the post-processed TNG300 simulation to make predictions directly comparable with observations.

\begin{center}
\begin{table*}
\centering
\begin{tabular}{cccc}
 Method &  ICL mass fraction (${\rm M_{200} \leq 10^{14} M_{\odot}}$) & ICL mass fraction  (${\rm M_{200} > 10^{14} M_{\odot}}$) & Total ICL mass fraction  \\
  &  (420 objects) & (280 objects) & (700 objects)  \\
\hline
\hline
\vspace{0.025cm}
 $r_{\rm ap}$ = 2\rhalf & $\ficl=0.32\pm 0.03$ & $\ficl= 0.33\pm 0.03$ & $\ficltot=0.16\pm 0.05$ \\ 
\vspace{0.025cm}
 $r_{\rm ap}$ = 5\% $R_{200}$ & $\ficl=0.45\pm 0.06$ & $\ficl= 0.50\pm 0.07$ & $\ficltot=0.22\pm 0.07$ \\
\vspace{0.025cm}
 $r_{\rm ap}$ = 10\% $R_{200}$ & $\ficl=0.29\pm 0.06$ & $\ficl= 0.34\pm 0.07$ & $\ficltot=0.15\pm 0.04$ \\ 
 \vspace{0.025cm}
 $r_{\rm ap}$ = 30 kpc & $\ficl=0.53\pm 0.06$ & $\ficl= 0.64\pm 0.08$ & $\ficltot=0.27\pm 0.08$ \\
 \vspace{0.025cm}
 $r_{\rm ap}$ = 50 kpc & $\ficl=0.41\pm 0.07$ & $\ficl= 0.54\pm 0.08$ & $\ficltot=0.22\pm 0.07$ \\
 \vspace{0.025cm}
 $r_{\rm ap}$ = 100 kpc & $\ficl=0.25\pm 0.06$ & $\ficl= 0.38\pm 0.09$ &  $\ficltot=0.14\pm 0.05$ \\
\hline
\hline
\vspace{0.1cm}
\end{tabular}
\caption{Means and standard deviation of different ICL fractions measured in this work using different aperture radii. While for $\ficl=\micl/(\micl+\mbcg)$ the means are presented in two mass ranges because most of these fractions vary with mass (see Fig.~\ref{fig:median_f_F_icl_vs_m200_diff_appertures}), for $\ficltot=\micl/(\micl+\mbcg+M_{\rm \ast,sat})$ the means are for the whole sample since these fractions weakly depend on mass. 
}
\label{tab:f_icl_our_results}
\end{table*}
\end{center}

\subsection{Comparisons with previous studies}
\subsubsection{The BCG+ICL mass fraction}
The most general comparison that we can make with previous works based on hydrodynamical simulations is with the stellar mass fraction in the BCG+ICL component, $F_{\rm BCG+ICL}$, without going into the difficulty of separating the ICL from the BCG. Since this fraction decreases with $\Mtwo$, in Fig.~\ref{fig:violin_plot} we showed the distributions of $F_{\rm BCG+ICL}$ in two cluster mass ranges. On average, clusters more massive than $\Mtwo=10^{14}~\msun$ exhibit $F_{\rm BCG+ICL}=0.55\pm 0.12$, similar to \citet{Pillepich2018a} (see their figure 10).
In \citet{Brough+2024} the authors analyzed 61 clusters in the $\Mtwo \approx 1-3 \times 10^{14}\Msun$ range from four different cosmological hydrodynamical simulations: Horizon-AGN \citep{dubois2014_Horizon-AGN}, Hydrangea \citep{bahe2017}, Magneticum\footnote{\url{www.magneticum.org/index.html}} and TNG100 \citep{nelson2019illustristng}. The overall mean fraction measured within 1 Mpc they report is $F_{\rm BCG+ICL}=0.58\pm 0.14$. Although three simulations agree with each other and with us (with Horizon-AGN having the lower fractions), the Magneticum clusters have significantly higher fractions, $F_{\rm BCG+ICL}=0.75\pm 0.10$, and correspondingly lower total satellite mass fractions. \citet{Brough+2024} interpret this as a result of their selection of very-relaxed systems from this simulation.  In our sample of clusters more massive than $\sim 10^{14}$ \msun\ there are a few ones with BCG+ICL mass fractions higher than 0.7, and these are the ones with the highest $\ctwo$ and $z_{50}$ values (Figs. \ref{fig:scatter_f_bcg_icl_vs_m200-c200_cc200-cz50}-\ref{fig:scatter_f_bcg_icl_vs_z50-deltam4_cdeltam4-cz50}), and thus the most relaxed. 

For 11 clusters in the $\Mtwo=1-7\times10^{14}$ \msun\ range, using the Horizon-AGN simulation, \citet{brown2024} report median and 16--84th percentiles for $F_{\rm BCG+ICL}$ of 0.45 and 0.40--0.49, respectively. Their slightly smaller fractions than ours with the TNG300 simulation is likely due to the topological method for determining substructures (the outputs from these simulations were calculated with the \textsc{AdaptHOP} halo finder, \citealp{Tweed+2009}), without employing an unbinding procedure, which may lead to the inclusion of more stellar particles in satellite galaxies, even if they are not dynamically bound \citep[see also][]{Brough+2024}.
\citet{contreras2024} analyzed 324 clusters more massive than $\sim 10^{15}$ \msun\ from the \textsc{The Three Hundred} simulation project with two different subgrid physics implementations, \textsc{Gadget-X} and \textsc{Gizmo-Simba}. The median and 16--84th percentiles fractions they find for \textsc{Gadget-X} are 0.51 and 0.36–0.64, while for \textsc{Gizmo-Simba} are 0.46 and 0.33–0.57, respectively. In our case, for the few clusters around $\sim 10^{15}$ \msun, $F_{\rm BCG+ICL}\approx 0.45$, in good agreement with these authors, taking into account that they measured the BCG+ICL mass fractions within $\Rfive$ (for $\Rtwo$, the fractions are slightly smaller).

\subsubsection{The ICL mass fractions}
\label{sec:ICL-comparisons}
As seen above, there is generally good agreement in the BCG+ICL mass fractions of the clusters obtained and measured in different cosmological hydrodynamical simulations, with the exception of Magneticum which produces much higher fractions (but this is apparently because the selected clusters are only relaxed). The comparison of the ICL fractions is more complicated because different authors use different criteria to separate the ICL from the BCG and because the reported fractions refer indistinctly to $\ficl=\micl/(\micl+\mbcg)$ or $\ficltot=\micl/(\micl+\mbcg+M_{\rm \ast,sat})$; see Table \ref{tab: Definition components} for definitions. Here, we have studied both fractions and reported results for different aperture radii, $r_{\rm ap}$, to separate the ICL from the BCG (subsects. \ref{sec:radial-distribution} and \ref{sec:ICL-fractions}). In Table \ref{tab:f_icl_our_results} we summarize our measurements, presenting the mean and standard deviations for $\ficl$ and $\ficltot$ using different $r_{\rm ap}$. As shown in Fig.~\ref{fig:median_f_F_icl_vs_m200_diff_appertures}, $\ficl$ may or may not correlate with $\Mtwo$, depending on the used $r_{\rm ap}$, while for $\ficltot$, the values are generally less dependent on $\Mtwo$, albeit with large scatter. 
Therefore, in Table \ref{tab:f_icl_our_results} the means for $\ficl$ are reported in two mass ranges, while for $\ficltot$, correspond to the whole sample.
Comparing our measurements with those of \citet[][their Figure 10]{Pillepich2018a}, who also analysed TNG300, we obtain slightly lower values for $\ficl$ and $\ficltot$ for both $r_{\rm ap}$=30 and 100 kpc (note that these authors measure the fractions up to $\Rfive$), while for $r_{\rm ap}$=2\rhalf\ the agreement is excellent. 

For the 61 clusters from the four cosmological simulations mentioned above, \citet{Brough+2024} measure $\ficltot$ for $r_{\rm ap}=$30 and 100 kpc (and up to 1 Mpc). The mean and standard deviations are $0.38\pm0.16$ and $0.22\pm 0.09$, respectively (see their Table 1). The very relaxed Magneticum clusters have ICL total mass fractions that are much larger than the other simulations, showing that their high $F_{\rm BCG+ICL}$ fractions are actually due to the high ICL contribution in these relaxed clusters. Our values of $\ficltot=0.27\pm0.08$ for $r_{\rm ap}=$30 kpc and $\ficltot=0.14\pm0.05$ for $r_{\rm ap}=$100 kpc agree very well with those measured for Hydrangea and TNG100, are lower than those for Horizon-AGN and much lower than those for Magneticum. 
In the case of the massive clusters from the \textsc{The Three Hundred} simulation project, for $r_{\rm ap}=50$ kpc (and up to $\Rtwo$), \citet{contreras2024} find median values of  $\ficltot\approx 0.30$ and $\approx 0.26$ for the  \textsc{Gadget-X} and \textsc{Gizmo-Simba} runs, respectively. For our clusters, $\ficltot\approx 0.22$ for $r_{\rm ap}=50$ kpc, lower on average than for the \textsc{The Three Hundred} simulations, but well within the dispersion, especially for the \textsc{Gizmo-Simba} runs.

For the 11 clusters selected from the Horizon-AGN simulation, \citet{brown2024} measure the total mass fractions in both the ICL and BCG components, but they do not use an aperture radius for this. They identified the ICL using the output of the \textsc{AdaptaHOP} halo finder as those stellar particles within the cluster that are not classified as part of any stellar structure.  The median and 16--84th percentiles of $\ficltot$ are 0.13 and 0.11--0.14, respectively. Interestingly, these fractions are only slightly lower than ours for $r_{\rm ap}=$100 kpc, 0.1$\Rtwo$ or 2\rhalf. From their Table 1 we also calculate the $\ficl$ fractions and find a mean of $\approx 0.30\pm0.07$ in agreement with our results for $r_{\rm ap}=$2\rhalf. 
For EAGLE, Cluster-EAGLE, and high-resolution 50 Mpc EAGLE simulations,  \citet{proctor2024} use a kinematical method to separate bulge, disc, and intrahalo stellar components.  They find $\ficl$ values around 0.3--0.4 for $\Mtwo\gtrsim 5\times 10^{13}$ \msun, which are in good agreement with our measurements for $r_{\rm ap}=$2\rhalf, $\ficl=0.33\pm0.03$.  
Finally, for the clusters in the Horizon Run 5 simulation, \citet{Joo2024} by using dynamical criteria to determine particles bound to density peaks (galaxies), have defined ICL particles as those not bound to any galaxy. At $z=0.625$, they obtain for $\sim 1200$ clusters values of $\ficltot\approx 0.12$, slightly increasing with time, such that if the simulation is continued, these values will likely be larger on average and consistent with our values of $\ficltot\approx 0.16\pm 0.05$ for $r_{\rm ap}=$2\rhalf. 
  
We conclude that in general, when adequately compared, the TNG300 clusters and the ones from other simulations show mass fractions of non-satellite stellar particles outside a given radius $r_{\rm ap}$ (the ICL mass or ICL total mass fractions) consistent among them. Furthermore, we have seen that measurements of the ICL fractions based on structure finders (both topological and kinematical) show results that agree with our measurements using $r_{\rm ap}=$2\rhalf. This supports our discussion in Sect. \ref{sec:optimal-radius} that 2\rhalf\ seems to be the radius where the gravitational influence of the BCG+inner dark matter halo ends, with stars freely floating within the cluster gravitational potential beginning to dominate the stellar density profile, i.e., 2\rhalf\ seems to be a suitable radius to nominally separate the ICL from the BCG.

\section{Summary and Conclusions}
\label{sec:conclusions}
We have used the 700 most massive structures, with virial masses from $\log(\Mtwo/\msun) \approx 13.7$
to 15.2, from the TNG-300 hydrodynamical simulation to study the (spherically averaged) radial distribution of stellar mass of their central objects, related to the BCG+ICL cluster component. Having explored the cluster properties on which the fraction $F_{\rm \ast,cen}=F_{\rm BCG+ICL}\equiv M_{\rm BCG+ICL}/M_{\rm \ast,200}$ may depend, we nominally separate the ICL from the BCG by introducing different aperture (transition) radii and study the ICL fractions obtained as a function of $\Mtwo$. We report values for two definitions of the ICL fraction: $\ficl=M_{\rm ICL}/M_{\rm BCG+ICL}$ and $\ficltot=M_{\rm ICL}/M_{\rm \ast,200}.$(for the definitions, see Table \ref{tab: Definition components}).  
For an aperture radius equal to 2\rhalf\, we have studied the distributions of $\ficl$ and $\ficltot$ and their dependences on some cluster properties and their dynamical state, as well as the contributions from different formation channels to the ICL and BCG components. 
The main results obtained from our study are as follows: 
\begin{itemize}

\item The BCG+ICL stellar mass correlates with the total cluster mass typically inferred from observations ($\Mfive$) as $M_{\rm BCG+ICL}\propto \Mfive^{0.76}$, with relaxed/intermediate clusters having a normalisation $\approx 32\%$ higher than disturbed. The median $F_{\rm BCG+ICL}$ values for the low ($13.7<\log(\Mtwo/\msun)\leq 14)$ and high ($\log(\Mtwo/\msun)>14$) mass clusters are 0.64 and 0.56, respectively, with a broad distribution skewed towards lower values (Fig.~\ref{fig:violin_plot}). 

\item $F_{\rm BCG+ICL}$ decreases with $\Mtwo$, but with a large scatter. Instead, we find that $F_{\rm BCG+ICL}$ is strongly correlated with both the NFW concentration $\ctwo$ and the redshift at which half the cluster mass has assembled, $z_{50}$ (Figs. \ref{fig:scatter_f_bcg_icl_vs_m200-c200_cc200-cz50}--\ref{fig:scatter_f_bcg_icl_vs_z50-deltam4_cdeltam4-cz50}). 
Moreover, for a given $\ctwo$, $F_{\rm BCG+ICL}$ is higher when $z_{50}$ is higher.  
We also found that $F_{\rm BCG+ICL}$ strongly correlates with the mass gap (e.g. between the most massive and the fourth most massive member galaxy, $\deltam$), a correlation that may eventually be compared with observational measurements. 

\item The stellar mass radial cumulative profile of the central object normalized to this mass at $\Rtwo$, $M_{\rm \ast,cen}$, follows a sigmoidal distribution as reported in \citet[][see our Fig.~\ref{fig:median_profile_f_bcg_icl_kpc}]{Pillepich2018a}, with the distribution being shallower the higher $\Mtwo$ is. This distribution, when expressed as a function of \rhalf\ is scale-independent and closely self-similar. By defining an aperture radius $r_{\rm ap}$ that separates the BCG from the ICL, we can calculate the BCG and ICL mass fractions, $f_{\rm BCG}$ and $\ficl=1-f_{\rm BCG}$, respectively, from the normalized cumulative profile.  
\item Different definitions of $r_{\rm ap}$ used in the literature give very different values for $\ficl$ and $\ficltot$, with different dispersion and dependences on $\Mtwo$, see Fig.~\ref{fig:median_f_F_icl_vs_m200_diff_appertures}.  For $r_{\rm ap}=30$, 50 and 100 kpc, $\ficl\propto\Mtwo^b$ with $b=0.22$, 0.26 and 0.29, and with a normalization that significantly decreases as  $r_{\rm ap}$ is larger. Regarding $\ficltot$, it is almost independent of $\Mtwo$ but with a large dispersion.  
For $r_{\rm ap}=0.05$ and 0.1$\Rtwo$, although it is scale dependent, $\ficl$ correlates still with mass with $b=0.17$ for both cases, while  $\ficltot$ is almost also independent of $\Mtwo$ and with a large dispersion.
For  $r_{\rm ap}=$ 2\rhalf, $\ficl$ is independent of $\Mtwo$ and nearly constant, with a mean of $0.33\pm 0.03$, while $\ficltot$ decreases slightly with $\Mtwo$, with a small dispersion. This is because $\ficltot=\ficl\times (1-F_{\rm \ast,sat}$), with $\ficl\approx$const. and  the mass fraction in satellites increasing with $\Mtwo$ (see Fig.~\ref{fig:scatter_f_bcg_icl_vs_m200-c200_cc200-cz50}). 

\item The radius 2\rhalf\ could be a reliable criterion to \textit{nominally} separate the ICL from the BCG, since it can be regarded as a transitional radius where the gravitational influence of the BCG+inner dark matter halo ends. Indeed, we find that  2\rhalf\ correlates with the NFW cluster scale radius, $r_s$, with 2\rhalf$\lesssim r_s$ (Fig.~\ref{fig:scatter_rhalf_m200_cc200_cm200}). The escape velocity within $\sim r_s$ is high so most of the stars that are tidally stripped or accreted in inner major mergers, cannot escape to larger radii despite their relatively high velocities. 

\item  While $\ficl$ has a very narrow normal distribution around 0.33, independent of any cluster property, $\ficltot$ has a broader distribution that segregates towards higher values for the more relaxed clusters (Fig.~\ref{fig:hist_f_F_ilc_dynamical_state}). Furthermore, $\ficltot$ is strongly correlated with $\ctwo$, $z_{\rm 50}$ and $\deltam$ (Fig.~\ref{fig:scatter_F_icl_vs_c200-deltam4_cz50_dyn_state}).

\item The in-situ star formation is more relevant in the growth of the BCG than the ICL: on average, $\sim 25$ percent of the BCG stellar masses and only  $\sim 8$ per cent of the ICL mass formed in situ (Fig.~\ref{fig:pie_charts_2samples_halo_mass}). For both components, the rest of their masses was accreted (ex situ). For the BCGs, the mergers, especially the major ones, are the main channel of ex-situ growth, while for the ICL, tidally stripped stellar particles from the surviving galaxies is more important, especially for the more massive clusters. 

\end{itemize}

Our results suggest that more concentrated clusters are likely to have stronger tidal fields, capable of stripping stellar particles from satellites or even disrupting them (mergers) in favour of the BCG and ICL components, and if the cluster was assembled earlier (higher $z_{50}$), these mechanisms act for longer, decreasing the mass fraction in the satellites $F_{\rm \ast,sat}$ in favour of the central object, i.e. increasing $F_{\rm BCG+ICL}$. However, there is no clear preference for the transfer of stellar particles to the BCG or the ICL, a process that appears to be more stochastic. The ex-situ growth of these components is more dominated by mergers, especially major, in the case of the BCG, and by tidal stripping of surviving satellites in the case of the ICL. In other words, the contribution of the different ex-situ channels (tidal stripping -which includes preprocessing-, and major, minor and very minor mergers) changes with the clustercentric radius. 

The strong correlation of the BCG+ICL or ICL mass fraction with $\ctwo$ could eventually be used to constrain this parameter for observed clusters. We have determined the best fit for $\ctwo$ given $F_{\rm BCG+ICL}$ (Eq. \ref{c200predict}), as well as an improvement of the fit introducing two more parameters related to the stellar distribution, $\ficl$ and $\deltam$ (Eq. \ref{c200predictwcor}). In the same spirit, the strong correlation between $\Mtwo$ and $\Mtwost$, could be used to constrain cluster masses from their observed total stellar masses (Eq. \ref{m200predict}).

We have shown that the ICL mass fractions determined here from the spherically averaged radial stellar mass profiles remain nearly the same for the projected distributions. Using the projected radius 2$r_{\rm \ast,half}^{2D}$ (which is $26\pm8$ percent smaller than 2\rhalf) to separate the ICL from the BCG in the projected mass distributions, $\ficlpro$ is  only $\sim 2\pm3$ per cent higher than $\ficl$, and $\ficltotpro$ is only $\sim 2\pm7$ per cent lower than $\ficltot$.  
Observers could use 2$r_{\rm \ast,half}^{2D}$ as the apperture radius to separate the ICL from the BCG, after masking or substracting satellites and other stellar features from the observed images. In any case, in order to obtain fair comparisons between simulations and observations, mass-to-light post-processing should be carried out and suitable simulated images obtained from the former. We will present this task in a future work.

\section*{Acknowledgements}
DMT thanks CONAHCyT for a PhD fellowship.
The authors acknowledge financial support from the CONAHCyT grant CF G-543 and the DGAPA-PAPIIT grants IN106823, IN111825 and IN108323. AM thanks DGAPA-UNAM for a postdoctoral fellowship. All the figures in
this paper were produced using the \textsc{\large matplotlib} \textsc{\large python} package \citep{Hunter2007}. This paper has been typeset with the Overleaf collaborative authoring tool\footnote{\url{https://www.overleaf.com/}}.

\section*{Data Availability}
The group catalogues and particle data for IllustrisTNG are publicly accessible at \url{https://www.tng-project.org/data/} and thoroughly described in \citet{tngdata}. The quantities computed in this work can
be obtained via a reasonable request to the corresponding author.



\bibliographystyle{mnras}
\bibliography{references.bib} 



\appendix
\section{Channels of BCG/ICL formation for different dynamical states}
\begin{figure*}
  \centering
    \includegraphics[width=\linewidth]{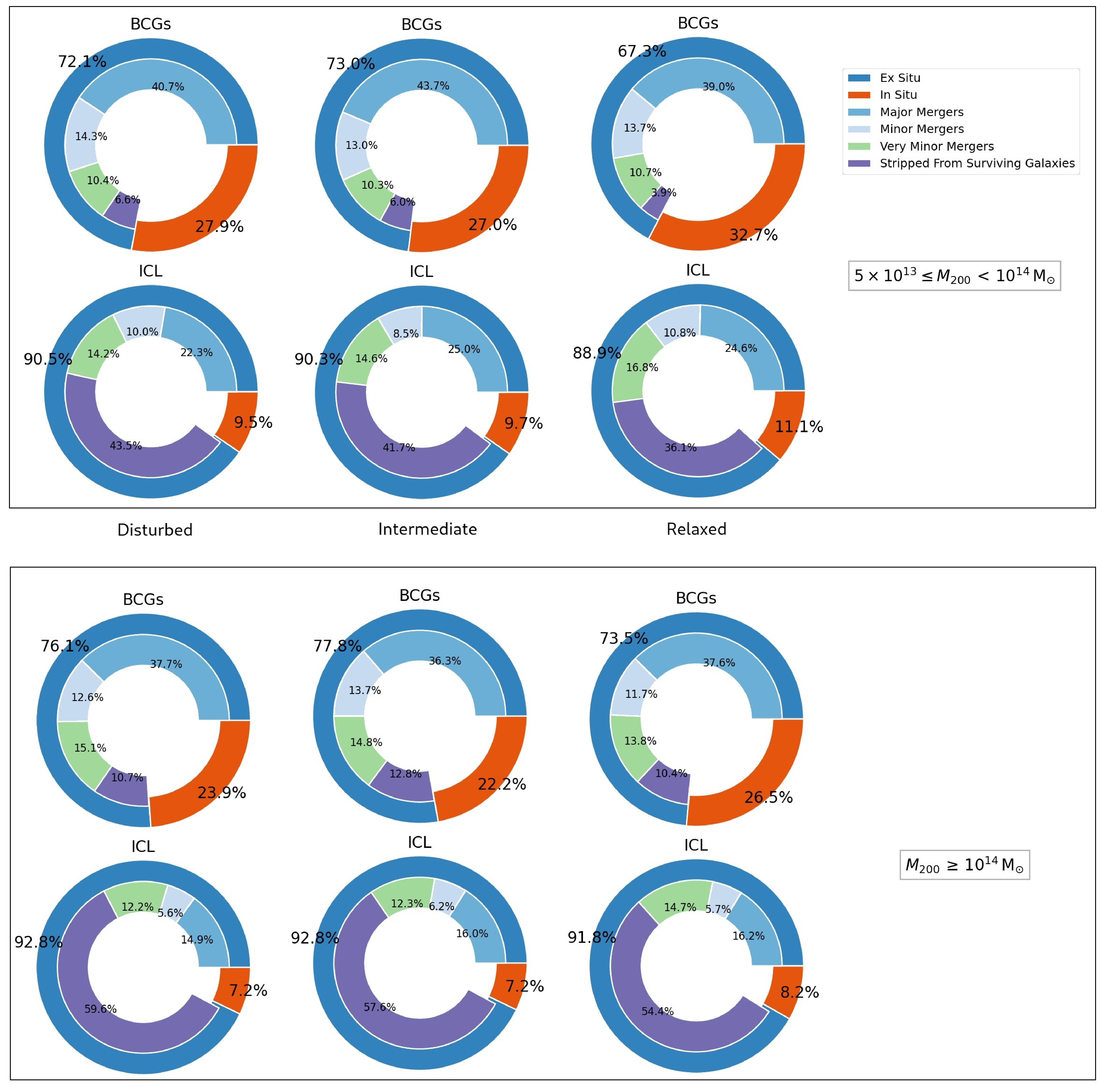}
	\caption{Mean stellar mass 'budget' of BCG (upper row) and ICL (lower row) for TNG300 clusters in the mass ranges $\Mtwo = 5 \times 10^{13} - 10^{14} \, \Msun$ (top panel) and $\Mtwo = 10^{14-15.2} \, \Msun$ (bottom panel).  
    The charter pies are as in Fig.~\ref{fig:pie_charts_2samples_halo_mass}, but separating the clusters into three dynamical states: disturbed, intermediate, and relaxed from left to right, respectively.
    }
	\label{fig:pie_charts_dyn_state_samples}
\end{figure*}

In Section~\ref{sec:chanels} we have presented the mean mass contributions of in-situ and ex-situ channels of BCG and ICL stellar mass growth, as well as the different sub-components of the ex-situ channel. Here, in Fig.~\ref{fig:pie_charts_dyn_state_samples} we present pie charts similar to those in Fig.~\ref{fig:pie_charts_2samples_halo_mass}, but for the three dynamical states in which we classified our clusters: disturbed, intermediate, and relaxed. In subsection \ref{sec:dynamical-states} we discuss these results.

\section{Posteriors for the parameters in the linear relation between $c_{200}$ and $F_{\rm BCG+ICL}$}\label{posterior}
In Section~\ref{proxy}, we have obtained a linear model for the relationship between $c_{200}$ and $F_{\rm BCG+ICL}$ (Eq.~\ref{c200predict}). The model contains three free parameters: slope ($a$), intercept ($b$), and intrinsic vertical scatter ($\sigma$). We derive the posterior distributions for these parameters through a Markov chain Monte Carlo (MCMC) sampling based on the log-likelihood in Eq.~(\ref{likelihood}) and a uniform prior\footnote{This is implemented using the \textsc{\large emcee} \textsc{\large python} package \citep{Foreman2013}}. For this, we generate four independent sample chains, each with a burn-in of 400 steps ($\approx 10$ times the auto-correlation time) and 4000 post burn-in steps. The final chain used to generate the posteriors is the concatenation of the four chains and exhibits strong convergence [the split $\hat{R}$ statistic \citep{Vehtari2021} is $\approx 1.00003$].

\begin{figure}
  \centering
    \includegraphics[width=0.99\linewidth]{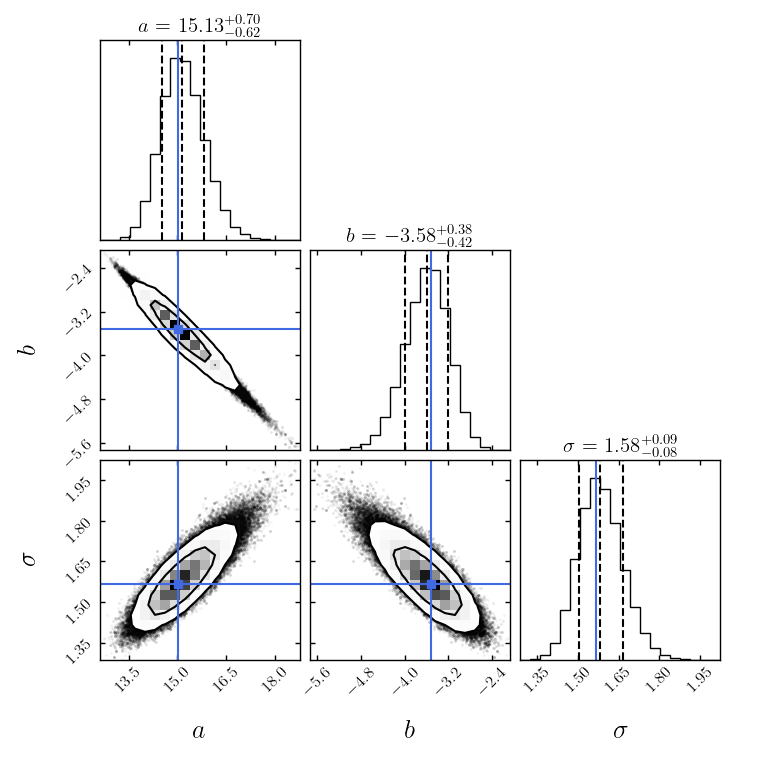}
	\caption{Posterior distributions for the free parameters in the linear fit to the $c_{200}$--$F_{\rm BCG+ICL}$ relation: slope ($a$), intercept ($b$), and intrinsic vertical scatter ($\sigma$). These have been generated based on the log-likelihood in equation~(\ref{likelihood})
    and under the assumption of a uniform prior. The solid blue line in each panel indicates the maximum likelihood solution (see Section~\ref{proxy}). The $x$-axis in each panel within a column is the parameter labelled beneath the bottom panel. The black dashed lines in the top-most panels show the 16th, 50th, and 84th percentiles for the parameter among the MCMC samples. The corresponding values
    for a given parameter are mentioned above the respective panel. The posteriors are well-behaved with Gaussian-like shapes. The maximum likelihood value for each parameter is always close to the peak of the derived posterior, and expected to vary by $\lesssim 9$ per cent based on the $1$-$\sigma$ interval.}
	\label{corner}
\end{figure}

The corresponding posteriors are illustrated using the corner plot\footnote{This is produced through the
\textsc{\large corner.py} code \citep{Foreman2016}.} in Fig.~\ref{corner}, where the top-most
panels show the distributions for the parameters, and the other panels show pair-wise correlations. The blue lines/points mark the maximum likelihood solution. The posteriors resemble Gaussian shapes, thereby implying a single maxima for all the parameters. The solution obtained by maximising the likelihood is expected to vary, at most, within the 16th and 84th percentiles of the posterior (dashed vertical lines). Based on these percentiles, our values can be considered to carry uncertainties of $\lesssim 6$ per cent for $a$, $\lesssim 9$ per cent for $b$, and $\lesssim 7$ per cent for $\sigma$. The solution is therefore well-constrained.

Furthermore, the $\sigma$ vs $a$ plot in the bottom-left panel suggests that, even though slopes significantly smaller than $a=15$ reduce the scatter, they deviate strongly from the most probable
value. To put it another way, if the $c_{200}$-$F_{\rm BCG+ICL}$ relationship is indeed linear, it is 
unlikely that it corresponds to the model with the least scatter -- like the one implied by least-squares.

Considering these results altogether, we conclude that the fitting parameters derived by us are sufficiently robust against statistical uncertainties, and represent the most-likely \textit{linear} description of the relationship between $c_{200}$ and $F_{\rm BCG+ICL}$.


\bsp	
\label{lastpage}
\end{document}